\documentclass[12pt]{iopart}
\usepackage[normalem]{ulem}
\usepackage{multirow}
\usepackage[utf8]{inputenc}
\usepackage{iopams}
\usepackage{hyperref}
\usepackage{color}  
\expandafter\let\csname equation*\endcsname\relax
\expandafter\let\csname endequation*\endcsname\relax
\usepackage{mathtools}
\usepackage[]{graphicx}
\usepackage{multicol}

\usepackage[multidot]{grffile}

\newcommand{\revisar}[1]{\textcolor{black}{#1}}
\newcommand{\nuevo}[1]{\textcolor{black}{#1}}

\graphicspath{{./}
	{figs/png/},
}
\DeclareGraphicsExtensions{.pdf, .eps, .png, .jpg, .gif}
\newcommand{\iotab}{\lower3pt\hbox{$\mathchar'26$}\mkern-7mu\iota}

\begin{document}

\title{Gyrokinetic simulations in stellarators using different computational domains}

\author{E. S\'anchez$^1$, J. M. Garc\'ia-Rega\~na$^1$, A. Bañón Navarro$^2$, J. H. E. Proll$^3$, C. Mora Moreno$^3$, A. González-Jerez$^1$, I. Calvo$^1$, R. Kleiber$^4$, J. Riemann$^4$, J. Smoniewski$^5$, M. Barnes$^6$, F. I. Parra$^6$}

\address{$^1$Laboratorio Nacional de Fusi\'on / CIEMAT. Avda Complutense 40, 28040, Madrid, Spain.}
\address{$^2$Max-Planck Insitut f\"ur Plasmaphysik, Garching, Germany.}
\address{$^3$Department of Applied Physics, Eindhoven University of Technology, Eindhoven, Netherlands.}
\address{$^4$Max-Planck Insitut f\"ur Plasmaphysik, Greifswald, Germany.}
\address{$^5$University of Wisconsin-Madison, 	Madison, Wisconsin, U.S.A.}
\address{$^6$Rudolf Peierls Centre for Theoretical Physics, University of Oxford, Oxford, United Kingdom.}
\ead{edi.sanchez@ciemat.es}
\vspace{10pt}

\begin{indented}
\item[]\today
\end{indented}

\begin{abstract}
	
In this work, we compare gyrokinetic simulations in stellarators using different computational domains, 
{namely, } flux tube,  full-flux-surface, and radially global {domains}.
Two problems are studied: the linear relaxation of zonal flows and the linear stability of ion temperature gradient (ITG) modes. Simulations are carried out with the codes EUTERPE, GENE, GENE-3D, and \texttt{stella} in magnetic configurations of LHD and W7-X using adiabatic electrons. The zonal flow relaxation properties 
obtained in different flux tubes are found to differ {with each other and with the radially global result, except for sufficiently long
	flux tubes, in general}. The flux tube length required for convergence is configuration-dependent. Similarly, for ITG instabilities, different flux tubes provide different results, {but the discrepancy between them diminishes with increasing flux tube length}. 
Full{-flux-}surface and flux tube simulations show good agreement in the calculation of the growth rate and frequency of the most unstable modes in LHD, \nuevo{while for W7-X differences in the growth rates are found between the flux tube and the full-flux-surface domains. Radially global simulations provide results close to the full-flux-surface ones}. \nuevo{The radial scale of unstable ITG modes is studied in global and flux tube simulations finding that  in W7-X, the radial scale of the most unstable modes depends on the binormal wavenumber, while in LHD no clear dependency is found}.

\end{abstract}
%
\vspace{2pc}

\section{INTRODUCTION}

    Gyrokinetics is the theoretical framework to study {micro}turbulence in magnetized plasmas. It takes advantage of scale separation between {turbulent fluctuations} and background quantities (such as magnetic geometry and plasma profiles), and provides a reduction of phase-space dimensionality, which allows an important saving of computational resources. In tokamaks, the theoretical analysis and the numerical simulation of microinstabilities and turbulence are largely facilitated by its axisymmetry, which makes all the field lines on a flux surface equivalent, so that simulations can be carried out in a reduced spatial domain called flux tube (FT): a volume extending several Larmor radii around a magnetic field line. Thanks to axisymmetry, the result of a calculation in a flux tube is independent of the chosen magnetic field line. Periodic boundary conditions in the parallel direction and the standard twist-and-shift formulation \cite{Beer95} are commonly used.

The lack of axisymmetry in stellarators introduces complexity at several levels. First, the twist-and-shift boundary condition in flux tube simulations is questionable \cite{Martin18} due to the three-dimensional dependence of the equilibrium quantities affecting micro-instabilities, such as magnetic field line curvature and magnetic shear. As a consequence of this dependence, different flux tubes over a given flux surface are in general not equivalent to each other \cite{Faber15}. A common practice when using flux tube codes in stellarators is to simulate the most unstable flux tube, which allows to quantify the upper bound of the instability.  However, the turbulence saturation level can be largely affected by the interaction between small-scale fluctuations and zonal flows (ZF), and the long time behaviour of the latter (which, in stellarators, shows distinct features as compared to tokamaks \cite{Mishchenko08,Helander11,Monreal17}) depends on the magnetic geometry of the whole flux surface. Different saturation mechanisms can dominate in different devices depending on the magnetic geometry \cite{Plunk17}. In addition, {the radial electric field, whose effect cannot be accounted for in a flux tube domain, might play a role in  the linear stability \cite{Villard02, Riemann16} and the turbulence saturation.}

In this work, we address the question of which is the minimum computational domain appropriate for the simulation of \revisar{instabilities and zonal flows} in stellarators and we study to what extent simplified setups, such as the flux tube approximation, can be used. For this purpose, we compare gyrokinetic simulations in different stellarator configurations using different computational domains and codes. The codes used are EUTERPE \cite{Jost01}, GENE \cite{Jenko00}, GENE-3D \cite{Maurer19} (the radially global version of GENE for stellarators), and \texttt{stella} \cite{Barnes19}, which cover different computational domains and implement different numerical methods. \texttt{stella} is a flux tube continuum code. Both a flux tube and a full-flux-surface version of GENE are available for stellarators. EUTERPE and GENE-3D are both radially global, although with different numerical schemes; EUTERPE is a particle-in-cell (PIC) code while GENE-3D is a {continuum} code.

Two problems are studied with these codes: the linear relaxation of zonal flows and the linear stability of Ion Temperature Gradient (ITG) modes. 
The first one has been studied both analytically and numerically in helical \cite{Sugama05,Sugama06,Sugama07} and general stellarator geometries   \cite{Mynick2007,Mishchenko08,Helander11,Monreal16,Monreal17},
showing distinct features as compared to tokamaks. The (semi)analytical calculations were validated against numerical simulations with EUTERPE, in the radially global (RG) domain, and GENE, in the full-flux-surface (FFS) domain, in W7-X and LHD configurations, finding a {remarkable} agreement \cite{Monreal16,Monreal17}. The linear relaxation of zonal flows in FT, FFS, and RG domains has been compared in HSX and NCSX configurations \cite{Smoniewski19}. 
In this work, we provide additional comparisons for LHD and W7-X in order to complete the picture of how the computational domain choice impacts the linear ZF relaxation. \revisar{We study this problem in the flux tube domain, and compare the results with those obtained in the FFS and RG domains, which was not previously done in these configurations.}
With respect to the {ITG} linear stability, no systematic comparison {between different computational domains has been reported so far} for stellarators. {In this contribution,} we compare results from ITG simulations carried out in FT, FFS and RG domains in the same configurations. Good convergence between different FT results is found in LHD,  while in W7-X significant differences are found between calculations in different flux tubes. 

The rest of the paper is organized as follows. In section \ref{secCodesAndDomains}, the computational domains and the codes used in this work are  introduced. {For LHD and W7-X and the computational domains considered, the collisionless linear ZF relaxation is studied in {section \ref{SecZFRelax}}, while the linear stability of the Ion Temperature Gradient (ITG) mode is presented in section \ref{SecLinearITGS}}. Finally, in section \ref{secSumandConcls} the results are discussed and some conclusions are {drawn}. 

		\section{Computational domains and codes}\label{secCodesAndDomains}

\begin{figure}
	\centering
	\includegraphics[trim=50 40 50 5, clip, width=8cm]{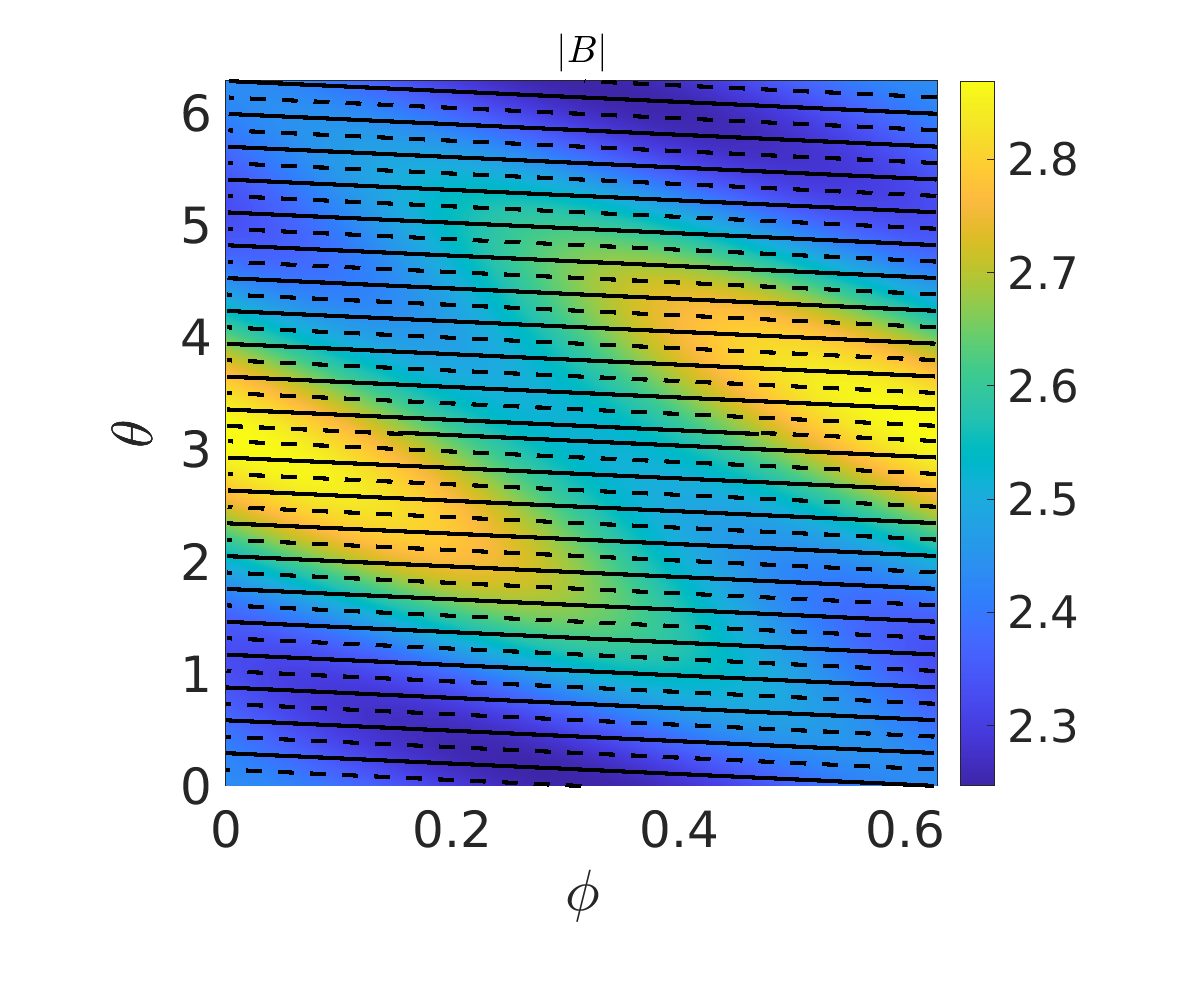}\\
	\includegraphics[trim=50 40 50 35, clip, width=8cm]{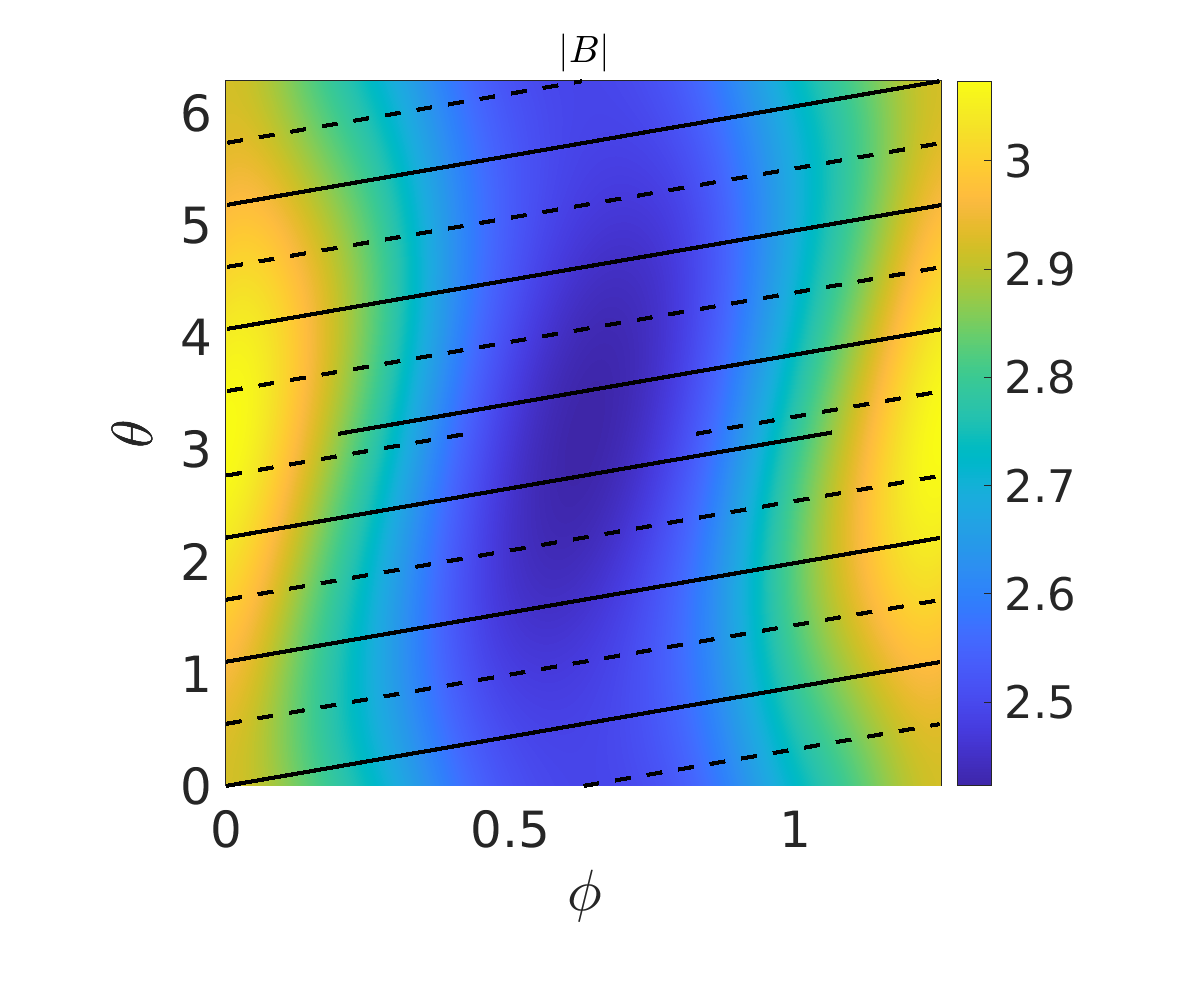}
	\caption{Magnetic field strength versus toroidal ($\phi$) and poloidal ($\theta$)  PEST angular coordinates for a  flux surface at middle radius, $r/a=0.5$, in a period of LHD (top) and W7X KJM (bottom) configurations. The field lines defining the two stellarator-symmetric flux tubes used in this work  (for the case $n_{pol}=1$) are shown. Thick line is used for the FT with $\alpha=0$ and dashed line for the one with $\alpha=\iota\pi/N$. The sections of the field lines laying in different device periods are all mapped to one period. }
	\label{FigMagFielQnttsOnePeriodNpol1}
\end{figure}

\begin{figure}
	\centering
	\includegraphics[trim=50 40 50 5, clip, width=8cm]{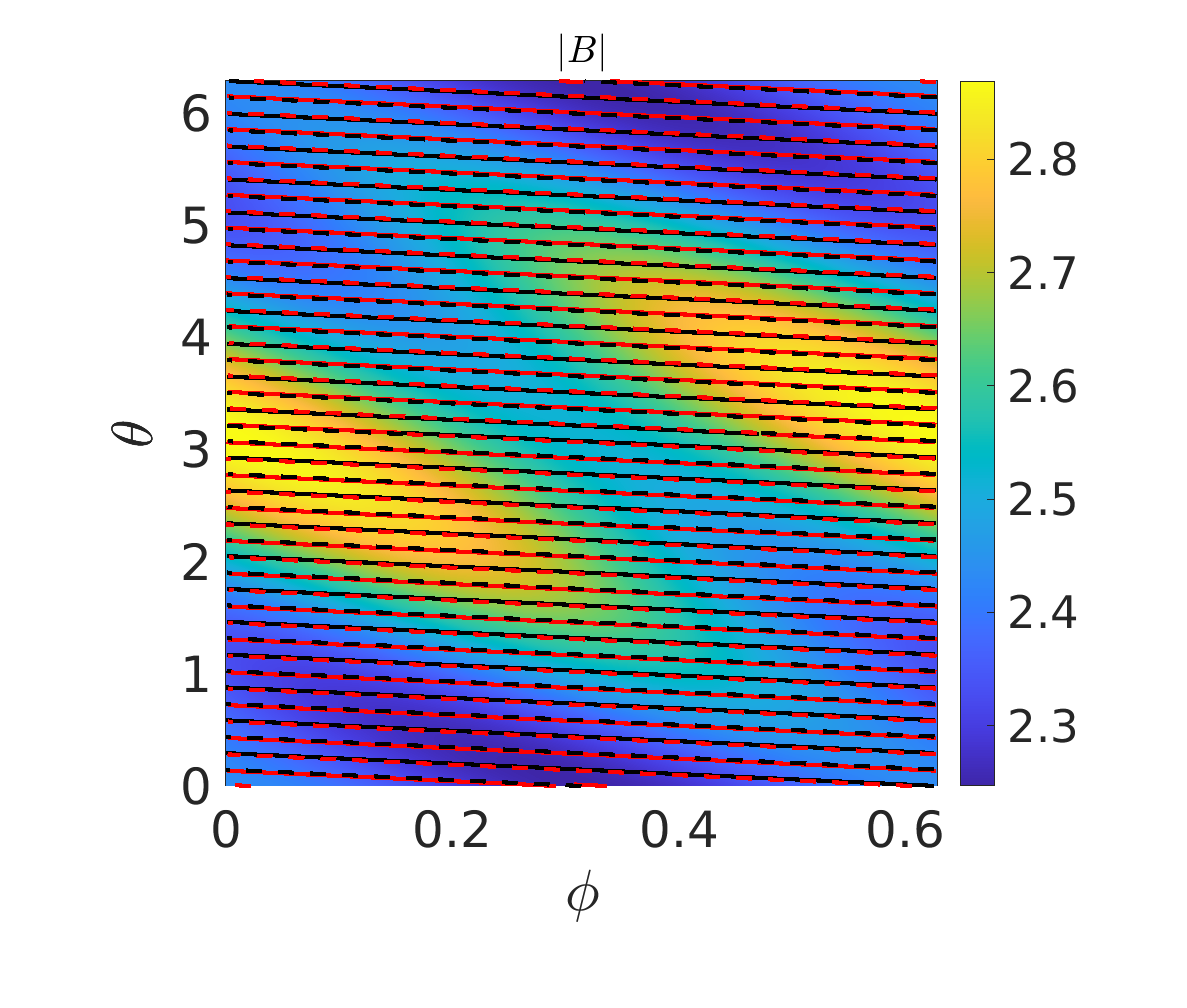}\\
	\includegraphics[trim=50 40 50 35, clip, width=8cm]{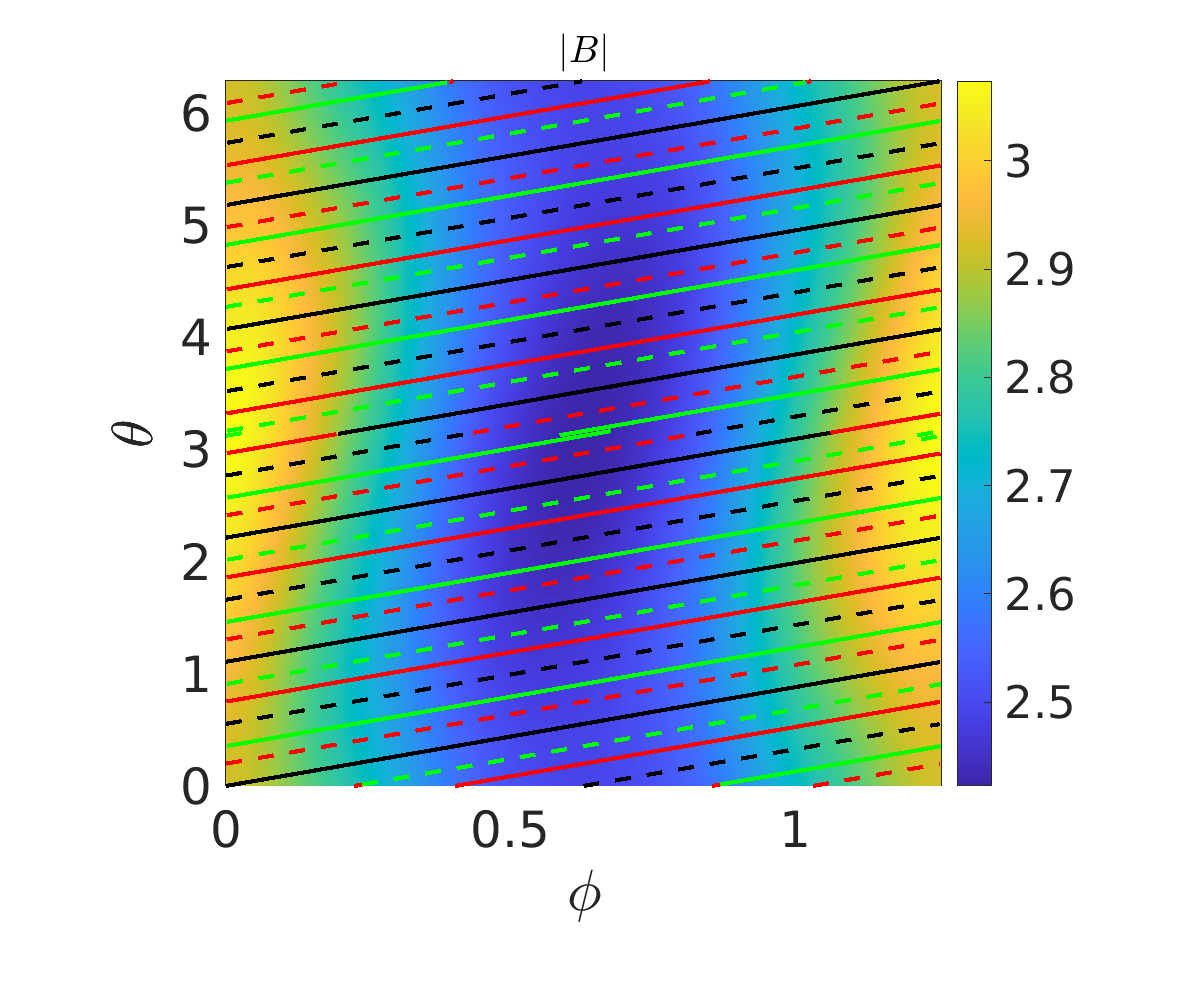}
	\caption{Magnetic field strength versus toroidal ($\phi$) and poloidal ($\theta$)  PEST angular coordinates for a  flux surface at middle radius, $r/a=0.5$,  in a period of LHD (top) and W7X KJM (bottom) configurations. The field lines defining the two stellarator-symmetric flux tubes used in this work are shown. Thick line is used for the FT with $\alpha=0$ and dashed line for the one with $\alpha=\iota\pi/N$. The sections of the field lines laying in different device periods are all mapped to one period. FTs of $n_{pol}=2$ and $n_{pol}=3$ are shown for LHD and W7-X, respectively. Black, red and green colors are used to show the first second and third poloidal turns, respectively. }
	\label{FigMagFielQnttsOnePeriod}
\end{figure}

\begin{figure*}
		\centering
		\includegraphics[trim=76 50 10 5, clip,width=17cm]{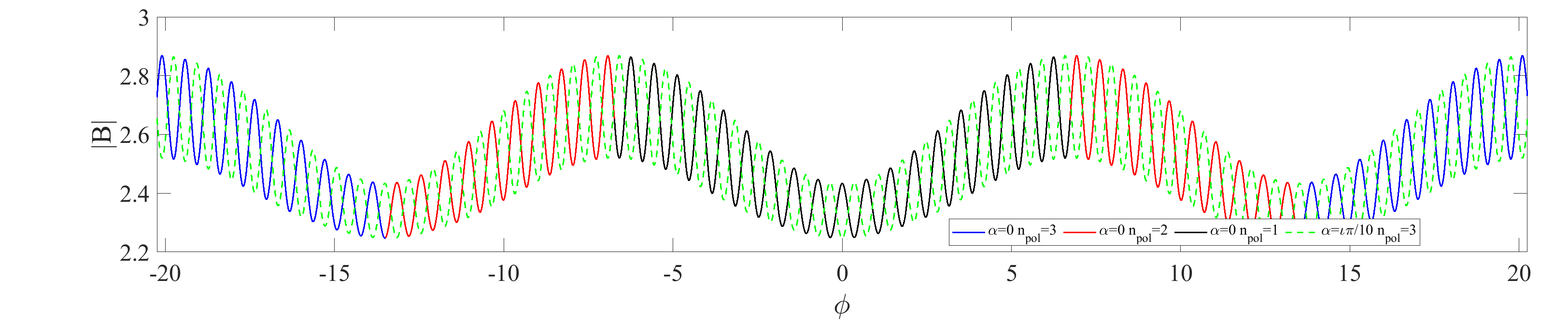}\\
		\includegraphics[trim=76 50 10 5, clip,width=17cm]{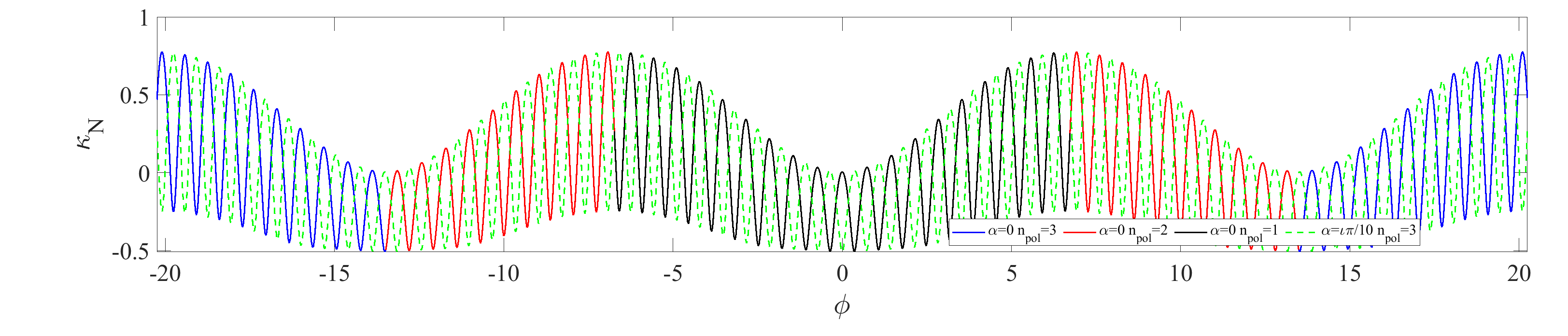}\\
		\includegraphics[trim=76 0 10 5, clip,width=17cm]{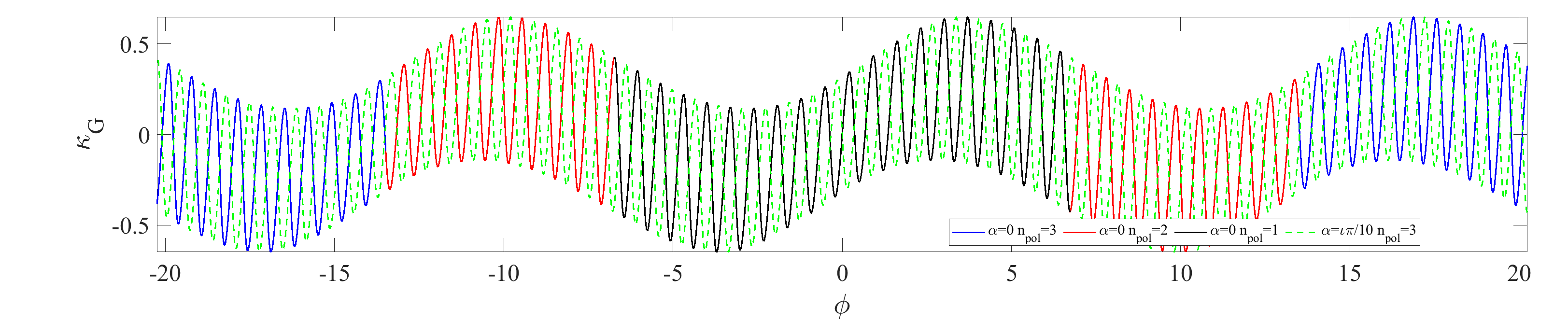}\\
		\includegraphics[trim=76 50 10 5, clip,width=17cm]{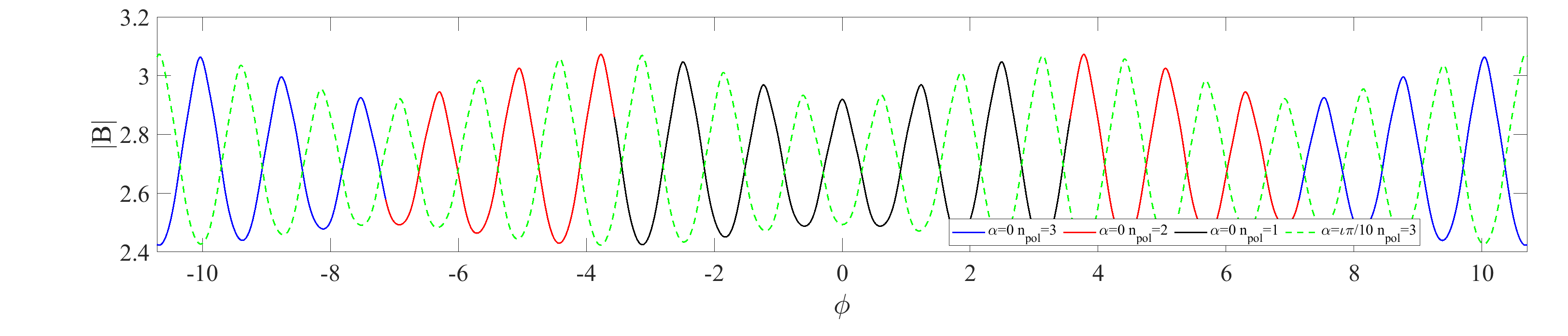}\\
		\includegraphics[trim=76 50 10 5, clip,width=17cm]{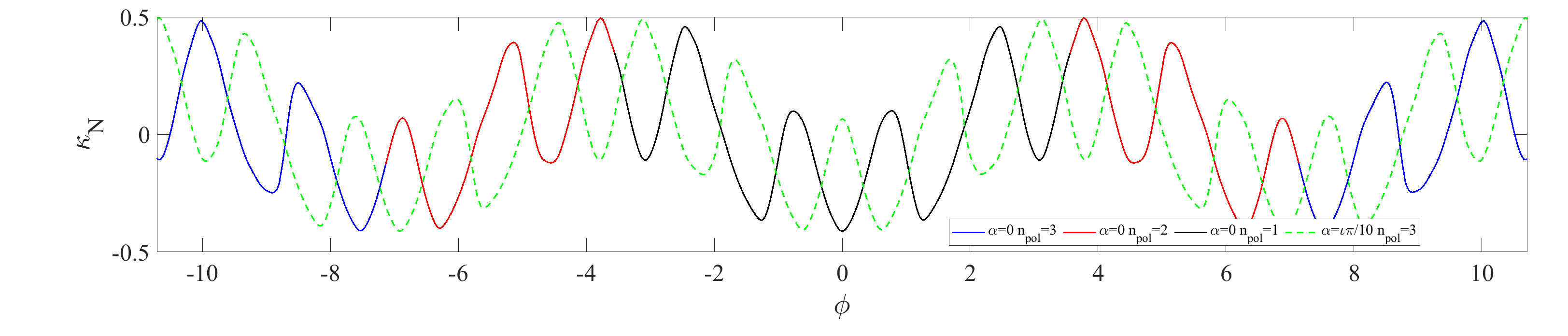}\\
		\includegraphics[trim=76 0 10 5, clip,width=17cm]{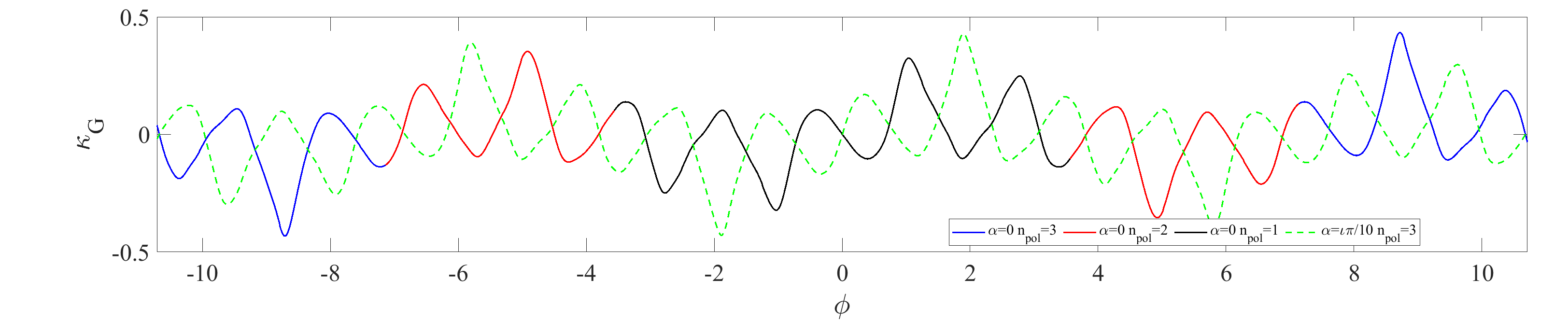}\\
		\caption{Magnetic field strength (rows 1 and 4), normal curvature $\kappa_N$ (rows 2 and 5) and geodesic curvature $\kappa_G$ (rows 3 and 6) along the field line in the $\alpha=0$ and $\alpha=2\pi/N$ flux tubes for a flux surface at middle radius ($r/a=0.5$) in the LHD (rows 1 to 3) and W7-X (rows 4 to 6) magnetic configurations. Different colors are used to highlight the extent covered by 1, 2 and 3 poloidal-turn FTs (only for the $\alpha=0$ FT). The horizontal axis shows the toroidal angle $\phi$ measured from the center of each flux tube.}
		\label{FigMagFielQntts}
	\end{figure*}

\nuevo{We devote this section to describing the different computational domains and codes used in this work, and start describing the simplest computational domain, the flux tube. A flux tube domain is defined by a volume around a magnetic field line, which extends a finite length along the direction parallel to the magnetic field. We will refer generically to  spatial coordinates $\{x,y,z\}$ that will be defined for each code later on, with $x$ a coordinate in the radial direction, $z$ a coordinate along the magnetic field line and $y$ a coordinate in the binormal direction, i.e. perpendicular to both the magnetic field and the flux surface normal vector. \revisar{Twist and shift }
	boundary conditions are defined in the direction parallel to the magnetic field \cite{Beer95}. The length of the flux tube is usually measured in terms of poloidal turns, $n_{pol}$. 
	 In a tokamak, the axisymmetry makes that a flux tube that extends one poloidal turn samples completely the geometry of a flux surface. In stellarators, this is not the case due to the three dimensional dependence of the magnetic geometry. 
	Flux tube simulations are local, i.e the width of the flux tube along the radial and binormal direction  is sufficiently thin so that the turbulence can be assumed to depend only on the local values of the relevant quantities of the problem (density, temperature, geometric coefficients, etc.) at the radial and binormal location of the selected line.  Periodic conditions in the radial direction are assumed. 
	 Only one radial mode $k_x$ and a perpendicular mode $k_y$ are scanned at a time in linear simulations.
	We consider two different flux tubes in this work: {one centered with respect to the position $(\theta,\zeta)=(0,0)$ and the other one centered with respect to the point} $(\theta,\zeta)=(0,\pi/N)$, with N the periodicity of the device (N=5 for W7-X and N=10 for LHD) and $\theta$ and $\zeta$ the angular PEST coordinates \cite{Grim83}. {Each of these flux tubes will eventually cover different lengths, which will be expressed in terms of the number of poloidal turns}. Figures \ref{FigMagFielQnttsOnePeriodNpol1} and \ref{FigMagFielQnttsOnePeriod}  show, for the mid plasma radius, the magnetic field strength in a period of LHD (top) and W7-X (bottom) as well as the flux tubes considered for different number of poloidal turns. 
	In Figure \ref{FigMagFielQntts}, several quantities related with the magnetic geometry that affect the linear zonal flow relaxation \cite{Monreal17,Sanchez15} and ITG instability are represented along these two FTs and for the same magnetic configurations. Both FTs are stellarator symmetric, which means that the magnetic field is the same at both ends of the FT. 
	A number of parallel wavenumbers $k_z$ depending on the spatial discretization along the field line is considered in the simulation. 	
	 In nonlinear simulations, a box in the radial and binormal directions is defined around the central field line defining the FT, and a set of radial and binormal wavenumbers $k_x$ and $k_y$ are considered. 
	 The fulfilment of the periodicity condition in the parallel direction for all the modes considered in the simulation is more restrictive in this case, with a strong influence of the magnetic shear \cite{Beer95, Martin18,Xanthopoulos09}.   	}

\nuevo{	The full-flux-surface computational domain is defined by a region around a specific magnetic flux surface, at a fixed value of the radial coordinate. In this computational domain the  locality along the binormal direction, which the FT approach assumes, is relaxed, while the domain is still local in the radial direction. This means that only the local values of magnetic field and profiles at the radial location at which the domain is centered are used in all the extent of the domain, and periodicity conditions are used in the radial direction. Multiple $k_z$ and $ k_y$ modes are considered in either a linear or nonlinear simulation.  Just a period of the device is simulated, usually, which means that $k_z$ modes that are not multiple of the device periodicity (N) are not considered in these simulations. In linear simulations either only one $k_x$ radial scale or a set of $k_x$ can be studied at a time.  In nonlinear simulations a box in the radial direction is defined around the flux surface and a range of radial wavenumbers $k_x$ is considered.  Similar to the FT case, some constraints are derived from setting periodicity conditions for all the $k_x$ considered. } 

\nuevo{The most complete computational domain is the radially global one. \revisar{This domain consist of a volume covering the full flux surface and  with a finite radial extent without assuming periodicity in the radial coordinate}. The number of modes considered in a simulation using this domain depends on the spatial grids used in the angular directions along the flux surfaces and the radial direction. As in the FFS domain, usually just a period of the device is simulated and the same limitations in respect to the toroidal Fourier modes apply to this case. The domain can cover either the full radius or an annular region with inner and outer boundaries different from the magnetic axis and the last closed flux surface. In this work we will refer to both kinds of simulations as radially global.  What distinguishes this domain from the FFS one is that in this case the values of magnetic field, the density and temperature profiles and their gradients are used at each specific position in the full 3D domain covered by the simulation. 
	The effects of global radial profiles, or large scale radial structures are captured in this domain, while they are not in the FFS one. Multiple modes in the radial direction are considered in this kind of simulation either linear or nonlinear, which are determined by the radial discretization. The boundary conditions used at the inner and outer boundaries of the domain are usually not fully realistic and can affect the simulation in a special way. It is common practice defining, either explicitly or implicitly, a buffer region close to each boundary in which the potential is assumed to be affected not only by physical aspects but also by the boundary conditions. The central radial region located far from the boundaries is considered not to be significantly affected by the boundary treatment.  
}

\nuevo{A radial electric field does not influence the result of flux tube simulations beyond a Doppler shift on the frequencies of the modes, while in FFS and RG domains, physical effects can be observed. The shearing effects of an electric field having a radial gradient cannot be easily accounted for in a FFS simulation domain while they are naturally included in a RG domain.}

The codes used for the simulations in this work are all based on the gyrokinetic formalism and solve the Vlasov-Poisson equations using the $\delta f$ approach, but they differ in several aspects of the numerical implementation.

EUTERPE is a particle-in-cell gyrokinetic code that can simulate {the entire confined plasma or a volume covering a radial annulus 
	 \cite{Jost01,Kornilov04}. For the {field solver}, natural boundary conditions are set at the inner boundary for the full {volume} simulations, while Dirichlet conditions are used {if the simulation is considering an annulus}. Dirichlet conditions are always used  at the outer boundary. Density and temperature profiles\nuevo{, depending only on the radial coordinate,} are used as input for both kinds of simulations. PEST ($s,\theta,\zeta$) magnetic coordinates \cite{Grim83} are used for the description of the fields, with $s={\Psi/\Psi_0}$, $\Psi$ the toroidal flux, $\Psi_0$ the toroidal flux at the {last close flux surface}, $\zeta$ the toroidal angle, and $\theta$ the poloidal angle. The electrostatic potential is Fourier transformed on each flux surface {considering a Fourier filter over a range of poloidal and toroidal mode numbers}. EUTERPE features the possibility of extracting a phase factor in linear simulations, thus allowing for a significant reduction in the computational resources \cite{Jost01}. This feature takes advantage of the translation property of the Fourier transform to center the Fourier filter on the region of interest in the wavenumber space and is used for the exploration of the wavenumber spectrum of unstable ITG modes in a wide range (see section 4). Even using these tools, the computational resources required for these simulations are significantly larger than for FT simulations. 
For details about the code EUTERPE, the reader is referred to \cite{Jost01,Kornilov04,Slaby18}.

The current version of the \texttt{stella} code \cite{Barnes19} solves, in the flux tube approximation, the gyrokinetic Vlasov and Poisson equations for an arbitrary number of species. The magnetic geometry can be specified either by the set of Miller’s parameters for a local tokamak equilibrium or by a three-dimensional equilibrium generated with VMEC \cite{Hirshman83}. The spatial coordinates that the code uses for stellarator simulations are{: the flux surface label $x=a\sqrt{s}$, with $a$ the minor radius; the magnetic field line label $y=a\sqrt{s_0}\alpha$, with $\alpha$ the Clebsch angle $\alpha=\theta-\iota\zeta$, $\iota$ the rotational transform and $s_0$ the radial location of the {magnetic field lines surrounded by the simulated flux tube}; and $z=\zeta$ the parallel coordinate}. The velocity coordinates are the magnetic moment $\mu$ and the parallel velocity $v_{\parallel}$.
\nuevo{A mixed implicit-explicit integration scheme, which can be varied from fully explicit to partial or fully implicit, is used in this code. 
	The implicit scheme has the advantage of allowing a larger time step than fully explicit approaches when considering kinetic electrons \cite{Barnes19}. }

GENE  is an Eulerian gyrokinetic code \cite{Jenko00}.  It uses a fixed grid in five-dimensional phase space (plus time), consisting of two velocity coordinates $v_{\parallel}$ (velocity parallel to the magnetic field) and $\mu$ (magnetic moment), and the three magnetic field-aligned coordinates $(x,y,z)$, with $x$ the radial coordinate, defined as $x=a\sqrt{s}$, $y$ the coordinate along the {binormal direction $\alpha$} and $z=\theta$ the coordinate along the field line. For stellarator simulations,  GENE can simulate either a flux tube or a full flux surface.  In the former option,  it is spectral in the coordinates perpendicular to the magnetic field $(x,y)$, 
while in the later it uses a representation in real space for the binormal coordinate $y$ and a Fourier representation for the radial coordinate. 
In both cases, periodic boundary conditions are used in the perpendicular directions.

GENE-3D is a global gyrokinetic code for stellarators \cite{Maurer19}. It solves the gyrokinetic equation in the same dimensional space as GENE.  However, it uses a real representation of the spatial coordinates. GENE-3D can be used in either FT, FFS or RG 
domains.  Periodic radial conditions are used when FT or FFS are simulated.  In the RG case, Dirichlet boundary conditions are used at both the inner and outer radial boundaries.

\revisar{In \texttt{stella}, GENE and GENE-3D it is common practice using hyperdiffusion terms, which allow the stabilization of sub-grid modes. While these terms are required for nonlinear simulations they can have a damping effect over the zonal components of the potential. In simulations devoted to study the linear relaxation of ZFs, like those presented in section \ref{SecZFRelax}, the hyperdiffusion has to be reduced significantly to avoid this damping, or alternatively, the resolution in velocity space has to be increased \cite{Pueschel14}.}

\nuevo{For all the codes, the} equilibrium magnetic field is obtained from a magneto-hydrodynamic equilibrium calculation with the code VMEC, and selected magnetic quantities are mapped for its use in the gyrokinetic codes with the intermediate programs VM2MAG, GIST and GVEC for EUTERPE, GENE and GENE-3D, respectively.  
\nuevo{Note that in \texttt{stella}, GENE and GENE-3D the resolution used in the equilibrium magnetic field is coupled to the resolution in the electrostatic potential, while in EUTERPE, both resolutions are independent. }

In the simulations presented in this work, the same density and temperature profiles are used in the global codes, using the local values of $n$, $T_e$, $T_i$, and their scale lengths  $L_{n}$, $L_{T_e}$, and $L_{T_i}$ (with  $L_X=|\frac{1}{X}\frac{dX}{dr}|$)  
at the reference  position, $r/a=0.5$, in the radially local codes,  with $r/a=\sqrt{s}$. 
\nuevo{The simulations presented here are all electrostatic and use adiabatic electrons.} 	

			\section{Linear relaxation of zonal flows}\label{SecZFRelax}

\begin{table*}[h]	
		\noindent
\footnotesize 
					\caption{	\label{tabZfs}Numerical settings used in the simulations of ZF relaxation in this paper. Radially local  simulations are run at $r/a=0.5$. RG simulations cover the full radius and the zonal electric field is extracted at $r/a=0.5$. $^\dagger$The resolutions in the MHD equilibrium magnetic field and the perturbed electrostatic field are independent in EUTERPE, while in \texttt{stella}, GENE and GENE-3D they are coupled. $^*$In EUTERPE, $n_x, n_y,n_z$ are the resolutions in the $s$, $\theta$, and $\phi$ PEST coordinates, respectively.}
					\noindent
			\begin{tabular}[h]{cp{1.2cm}ccccccccccc}
				\cr
				\multirow{4}{*}{{LHD}}	& code 		& \multicolumn{1}{c}{domain }   & 	& markers&$\Delta m,n$ & $n_x^*$ & $n_y^*$  & $n_z^*$ & $n_{v_{\par}}$ & $n_{\mu}$ & hyp'z & hyp'v \cr  
				\hline 
				\hline
				& \multirow{2}{*}{GENE} & {FT} & & & & 1 & 1 & $256 \times n_{pol}$ & 144 & 28 & 0.1 & 0.1 \cr
				\cline{4-12} 
				&  & {{FFS}} & & & & {1} & {1} & {$256$} & {144} & {28} & {0.1} & {0.1} \cr
				\cline{3-12}%
				& \multirow{2}{*}{EUTERPE}&\multirow{2}{*}{RG$^\dagger$} &${\varphi}$ & 50-240 M  & 6 	& 24-240	& 32 	& 32 	&  -	& -	& -	&  -\cr
				\cline{4-12}
				& 							& 							 & MHD 	& 	&  		& 			& 141	&  256 	& 256	&  \cr
				\hline
				\hline
				\cr
				\hline
				\hline
				\multirow{4}{*}{{W7-X}}	& {GENE} & { FT} & &&  &  1 & 1 & $128 \times n_{pol}$ & 256 & 32 & 0.01 &  0.01\cr 
				\cline{3-12}
				& \texttt{stella} & { FT} &&   & & 1 & 1 & $ 128 \times n_{pol}$ & 256 & 32 & - &  -\cr 
				\cline{3-12} 
				& \multirow{2}{*}{EUTERPE} &\multirow{2}{*}{RG$^\dagger$} & ${\varphi}$ & 50-500 M  & 6	& 24-600	& 32 & 32 &  -& -& -&  -\cr
				\cline{4-12}
				& & & MHD & & &  & 99	& 256 & 256 \cr								
				\hline									
				\hline		
			\end{tabular}
\end{table*}
  We start the comparison of computational domains {with} simulations of linear collisionless relaxation of ZFs. The long-term linear response of ZFs in stellararators is a paradigmatic problem for our purpose, i.e. the comparison of results in different computational domains, because it depends on quantities averaged over the full flux surface and exhibits distinct features in stellarators as compared to tokamaks \cite{Mynick2007,Mishchenko08,Helander11,Monreal17}. \nuevo{This problem has been studied numerically in different stellarator configurations in \cite{Sugama05,Sugama06,Sugama07,Kleiber10,Helander11,Monreal17,Sanchez13,Smoniewski19} and the low frequency oscillation of ZFs, specific of stellarators, was identified experimentally in TJ-II \cite{Alonso17,Sanchez18}.}
The relaxation of ZFs \nuevo{has already been studied {for the} quasisymmetric devices HSX and NCSX in the FT, FFS and RG domains in \cite{Smoniewski19} and in W7-X configurations in FFS and RG computational domains in \cite{Monreal16,Monreal17}}. Here we use the standard configuration of LHD, with $R_{ax} = 3.74 ~\rm{m}$ and $B_0 = 2.53 ~\rm{T}$ and a high mirror (KJM) configuration of W7-X, with beta 3\%. 
For the comparison of computational domains we use  simulations in FT geometry carried out with GENE \nuevo{and \texttt{stella}, in FFS  with GENE,} and RG 
carried out with EUTERPE.

For the simulations with EUTERPE, covering the full radial domain, flat density and temperature profiles are used with $T_i=T_e=5 ~\rm{keV}$ and $n=10^{19} ~\rm{m}^{-3}$ in both configurations. The simulations are initiated by setting an initial perturbation to the ion distribution function that is homogeneous on the flux surface and having a Maxwellian distribution in velocities. The radial dependence of the initial perturbation is such that a perturbation to the potential with the form $\varphi\propto \cos(k_s \pi s)$ is obtained after the first computation step. Semi-integer values are chosen for $k_s$, so that the initial perturbation is consistent with boundary conditions. With this definition, the perpendicular wavevector is $k_{\perp}=k_s\pi|\nabla s|$.
The simulations are evolved in time, the zonal electric field is {diagnosed at $r/a=0.5$} and its time trace 
is fit to a model\footnote{While for zonal perturbations with the general form $\varphi_k(s,t)e^{k_st}$ the residual level is the same for the potential and for its radial derivative (see \cite{Monreal16}), for EUTERPE simulations the second one is preferred because of practical reasons.}. 
\begin{equation}
\frac{\varphi'(t)}{\varphi'(0)} = A \cos(\Omega t) e^{-\gamma_{ZF} t} + \frac{c}{1+dt^e} + R,
\label{EqModelFit}
\end{equation}
from which the ZF residual level, $R$, and the oscillation frequency, $\Omega$, are extracted (see \cite{Monreal16,Mishchenko08,Helander11,Monreal17} for details about the calculation of the residual level and frequency in general stellarator geometry). The simulations are electrostatic and a long wavelength approximation in the quasineutrality equation is used, which is valid for modes with normalized radial wavelength {$k_{\perp}\rho_i < 1$},  with $\rho_i=\sqrt{2mT}/eB$, with $m$ the ion mass, $e$ the elementary charge, $T$ the local temperature and $B$ the magnetic field strength. This approximation limits the extraction of ZF relaxation properties for large radial wavenumbers. 
The numerical parameters for these simulations are as follows. The resolution in angular coordinates is $n_{\theta}= n_{\zeta}=32 $, with $n_{\theta}$ and $n_{\zeta}$ the number of points in the $\theta$ and $\zeta$ grids, respectively. \nuevo{
	 These resolutions correspond to those used for the perturbed electrostic potential.} Only 6 toroidal and poloidal modes are kept in the simulation; larger modes are suppressed by using a \nuevo{squared} Fourier filter. \nuevo{A diagonal filter suppressing modes with $|m-n\iotab| > 15$ is also used}. The radial resolution is set to a value that properly resolves the {potential} perturbation, and it is increased linearly with the radial wavenumber. As the radial resolution is increased, the number of markers is increased accordingly {in order to} keep the ratio of makers per mode constant. \nuevo{In this kind of simulations a large number of markers, as compared to linear ITG simulations, is required, particularly for large radial wavenumbers (see Table \ref{tabZfs}). A large number of markers in EUTERPE translates into a better resolution in velocity space. The resolutions used in all the simulations presented in this section are compiled in table \ref{tabZfs}.}.

Simulations in FT computational domain are carried out \nuevo{with both GENE and \texttt{stella}.} \nuevo{The simulations are initialized} by setting up an initial perturbation with radial scale determined by $k_x$ and symmetric along the 
binormal direction ($k_y=0$). 
The simulation is linearly evolved in time and the real component of the electrostatic potential is registered and fit using the same model (Eq. \ref{EqModelFit}) to extract the frequency and the residual level. For the comparison with global simulations we use the radial scale of the perturbation $k_x$ normalized with the reference Larmor radius, {$\rho_{r}$, with $\rho_r=\sqrt{2mT}/eB_r$}, $T$ the local temperature at the radial position of the simulation, and {$B_r$ the reference magnetic field.} The wavenumbers $k_x$, used in FT simulations, and $k_s$ used in RG simulations are related by $k_x=2 k_s\pi r/a^2$. \nuevo{It should be noted that in order to make a reliable fit of the potential time traces and obtain accurate values for $R$ and $\Omega$ using Eq. \ref{EqModelFit}, a long time trace is required, which is difficult, particularly for large radial wavenumbers. 
	Both, spatial and velocity-space resolutions and hyper-diffusivities have to be set to specific values in order to minimize the ZF damping. 
}
\nuevo{The ZF relaxation was also studied in FFS simulations carried out with GENE. Similar to the FT ones, the simulation is initiated by setting a perturbation in a specific radial scale $k_x$ and evolved linearly and without collisions. The time evolution of the zonal potential is fit using the same model.  FFS simulations were carried out in the LHD configuration only.   The very small magnetic shear of the W7-X configuration selected made very difficult getting reliable results in FFS simulations for finite $k_x$ values. The numerical settings for these simulations are listed in Table \ref{tabZfs}.}

The results for the residual level $R$ obtained by fitting the {time trace of the} zonal potential/electric field to the model (Eq. \ref{EqModelFit}) 
 for FT{, FFS} 
and RG domains  for LHD and W7-X configurations are compared in Figure \ref{FigZF1}. {The residual level increases} with the radial {wavenumber} of the perturbation in both LHD and W7-X configurations, in agreement with previous calculations \cite{Sugama06,Monreal16,Smoniewski19}. {However,} the residual level for a given radial scale {is} significantly larger in the case of W7-X {as compared with LHD}. The maximum {value of} the residual is also found for a larger wavenumber, close to $k_x\rho_r \sim 1$ in the W7-X.

In LHD, the agreement between calculations in different flux tubes, {considering} 1, 2 and 3 poloidal turns, is {excellent}. {For the $\alpha=0$ flux tube the shortest extension of $n_{pol}=1$ exhibits just slightly smaller residual levels than the other lengthier choices}. The agreement between the FT and FFS 
{results is also remarkable}. This is due to the fact that the geometry explored by flux tubes, even with just one poloidal turn, \nuevo{ is densely distributed over the flux surface (see Figure \ref{FigMagFielQnttsOnePeriodNpol1}). When the FT length is doubled  the location of the magnetic field lines along the  second poloidal turn almost overlaps the first one,  which explains the result that FTs with $n_{pol}=2$ provide results very similar to those of $n_{pol}>3$ (not shown in the figure), and also close to those with $n_{pol}=1$ (see Figures \ref{FigMagFielQnttsOnePeriod} and \ref{FigMagFielQntts})}. The global simulation matches the FT and FFS 
results better for small radial wavenumbers, while for {$k_x\rho_r>0.25$} the residual level is larger than those of the FT and FFS 
domains by {approximately} 7\% for LHD. 
\begin{figure}
	\centering
	\includegraphics[trim=20 67 70 10, clip, width=8.2cm]{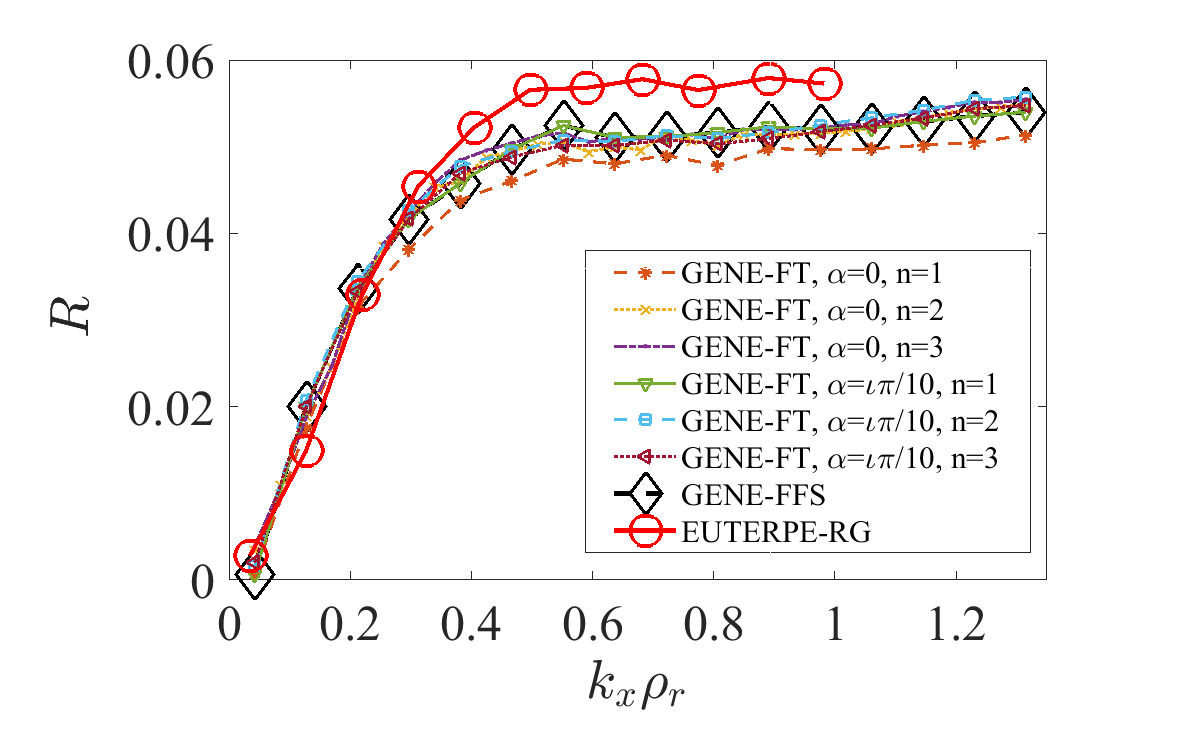}
	\includegraphics[trim=20 57 70 10, clip, width=8.2cm]{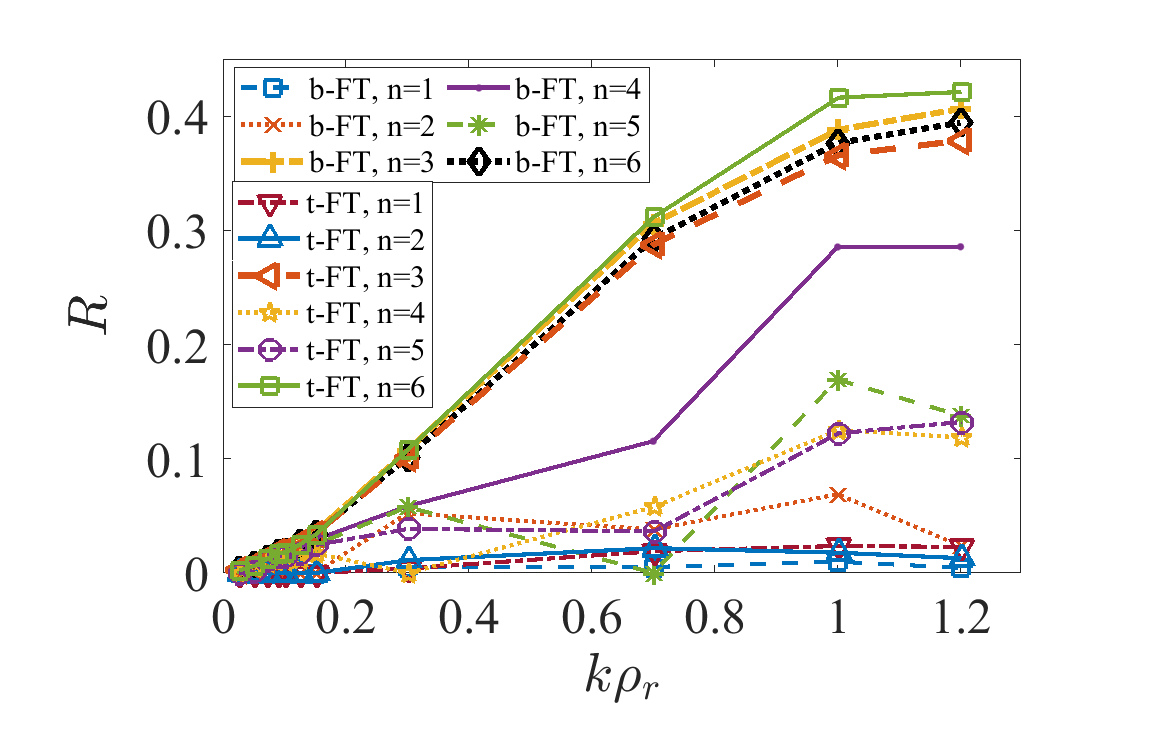}
	\includegraphics[trim=20 10 70 10,
			width=8.25cm]{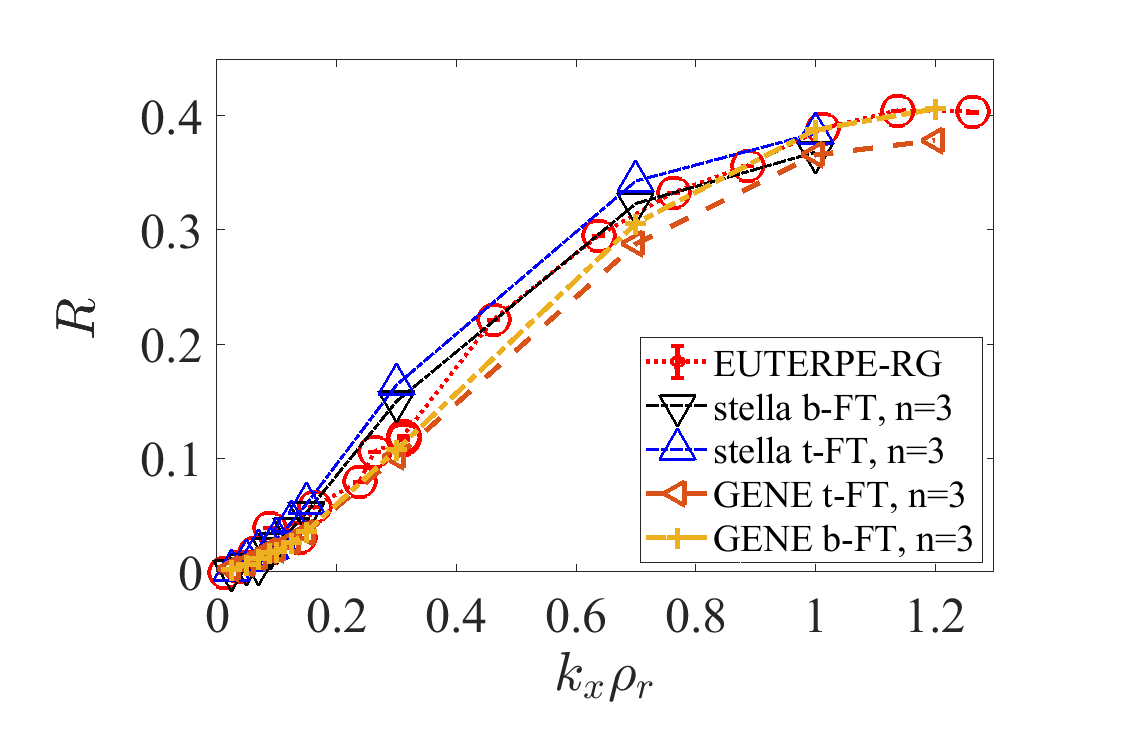}
	\caption{Long-term ZF residual level vs radial scale $k_x$ in the FT and FFS (GENE) and RG (EUTERPE) computational domains in LHD (top panel) and in FT and RG domains in W7-X (middle and bottom panels) at r/a=0.5. Different flux tube lengths $n_{pol}=1-4$ are considered. The bottom panel shows the residual level vs $k_x$ in W7-X for RG simulations with EUTERPE and FT simulations with \texttt{stella} and GENE with $n_{pol}=3$. The FT length is labeled $n=1,2...$ to shorten the legends. }
	\label{FigZF1}
\end{figure}

In the W7-X configuration, the situation  differs significantly. Different flux tubes provide different \nuevo{results, as  demonstrated in Figures \ref{FigZF1} and  \ref{FigZDepWithLength}. The middle pannel in Figure \ref{FigZF1} shows the residual level vs radial scale of the perturbation in the bean and triangular flux tubes (labeled ``b-FT" and ``t-FT" in the figures) in W7-X for different flux tube lengths. Figure \ref{FigZDepWithLength} shows the dependence of the residual level with the FT length for several radial wavenumbers $k_x=0.15,  0.3,  0.7,1$. From these figures it is clear, first, that different FTs provide different results for the ZF residual, in general, and second, that for a given FT different results  are obtained depending on the FT length.} 
\nuevo{	The explanation for this is related to the fact that regions of the flux surface with  different properties are explored by the flux tubes in each poloidal turn. 
	As shown in Figure \ref{FigMagFielQnttsOnePeriodNpol1}, for W7-X the flux surface is not as densely covered using $n_{pol}=1$ as in the case of LHD. }
	\nuevo{When the length of the FT is increased to $n_{pol}=2$ and $n_{pol}=3$ the flux surface is more densely covered and for  $n_{pol}=3$ the triangular FT crosses locations explored by the bean FT for  $n_{pol}=1$, as shown in Figure \ref{FigMagFielQnttsOnePeriod}. The residual level obtained {for the} shortest FTs is very small and increases {with the flux tube length}, thus getting close to the RG result for $n_{pol}=3$, in agreement with expectations from Figure \ref{FigMagFielQnttsOnePeriod}. For $n_{pol}=3$ the results from FT calculations both for the bean and triangular FTs with \texttt{stella} and GENE are in good agreement. Interestingly, the triangular flux tube seems to converge to the bean flux tube (and to the RG result) for $n_{pol}=3$ (see bottom panel of Figure \ref{FigZF1}). However, increasing the length further to $n_{pol}=4$ makes the results of bean and triangular FTs separate again.  Convergence between flux tubes and also to the RG results is reached again for $n_{pol}=6$ (see Figure \ref{FigZDepWithLength}).
}

\begin{figure}
	\centering
	\includegraphics[trim=5 35 60 10, clip, width=8.25cm]{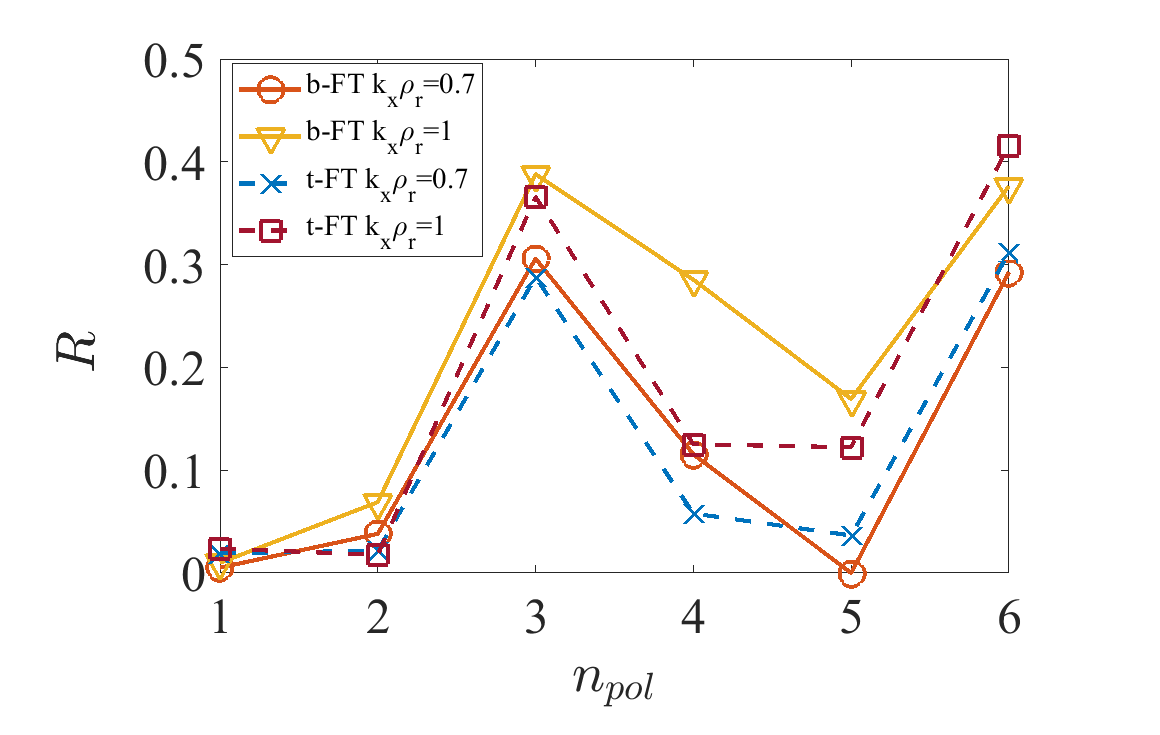}
	\includegraphics[trim=5 10 60 10, clip, width=8.25cm]{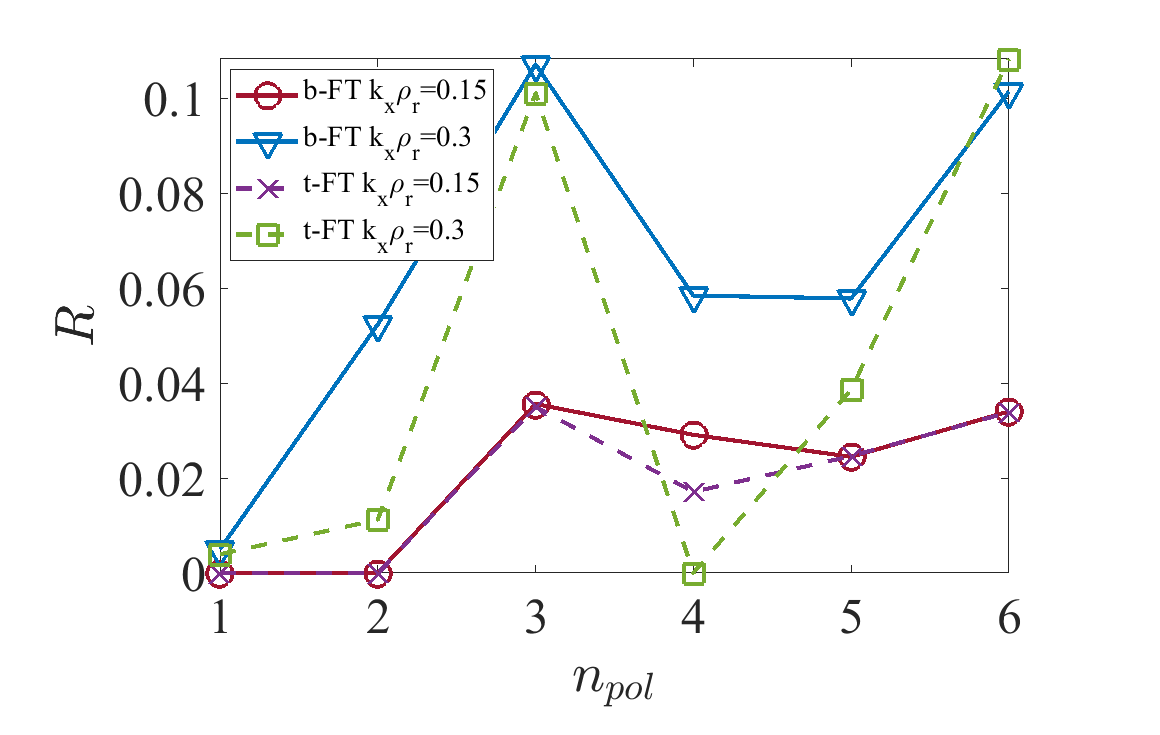}
	\caption{Dependency of the ZF residual level, R, with the FT length for the two FTs and several radial scales of the perturbation $k_x$, obtained from simulations with GENE in the W7-X KJM magnetic configuration.}
	\label{FigZDepWithLength}
\end{figure}
	
%
\begin{figure}
	\centering
\includegraphics[trim=20 10 70 10, clip,width=8.25cm]{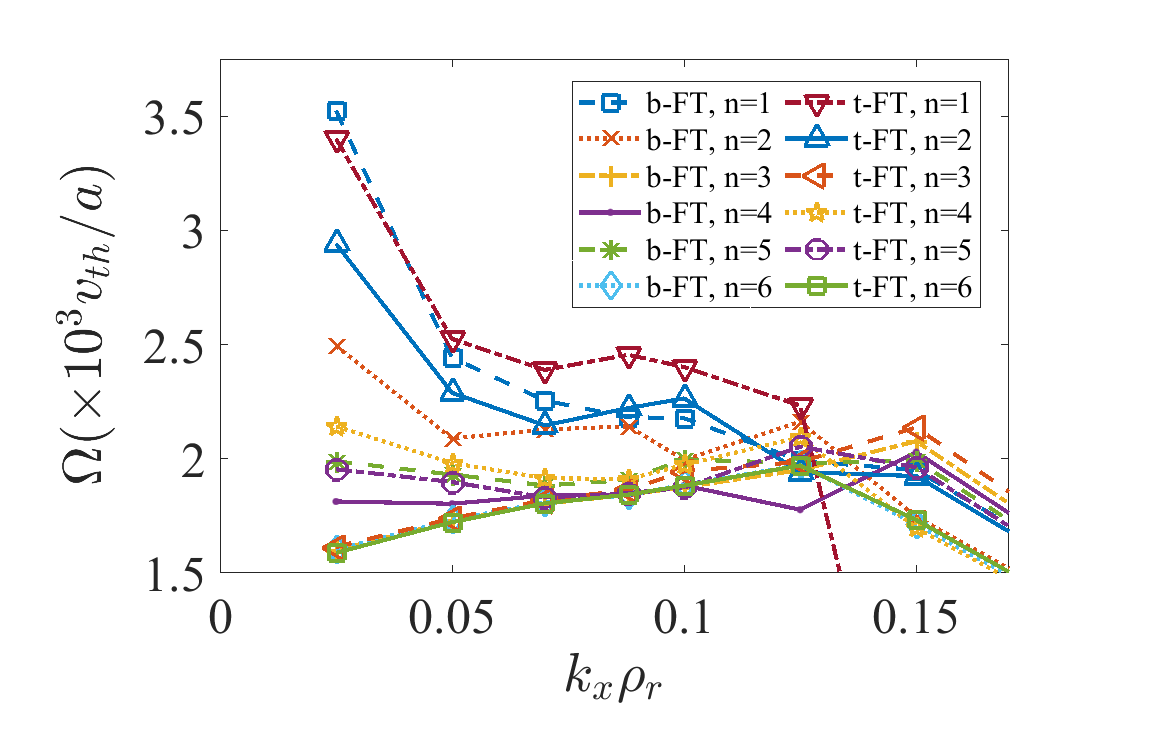}
	\includegraphics[trim=30 10 65 10, clip, width=8.25cm]{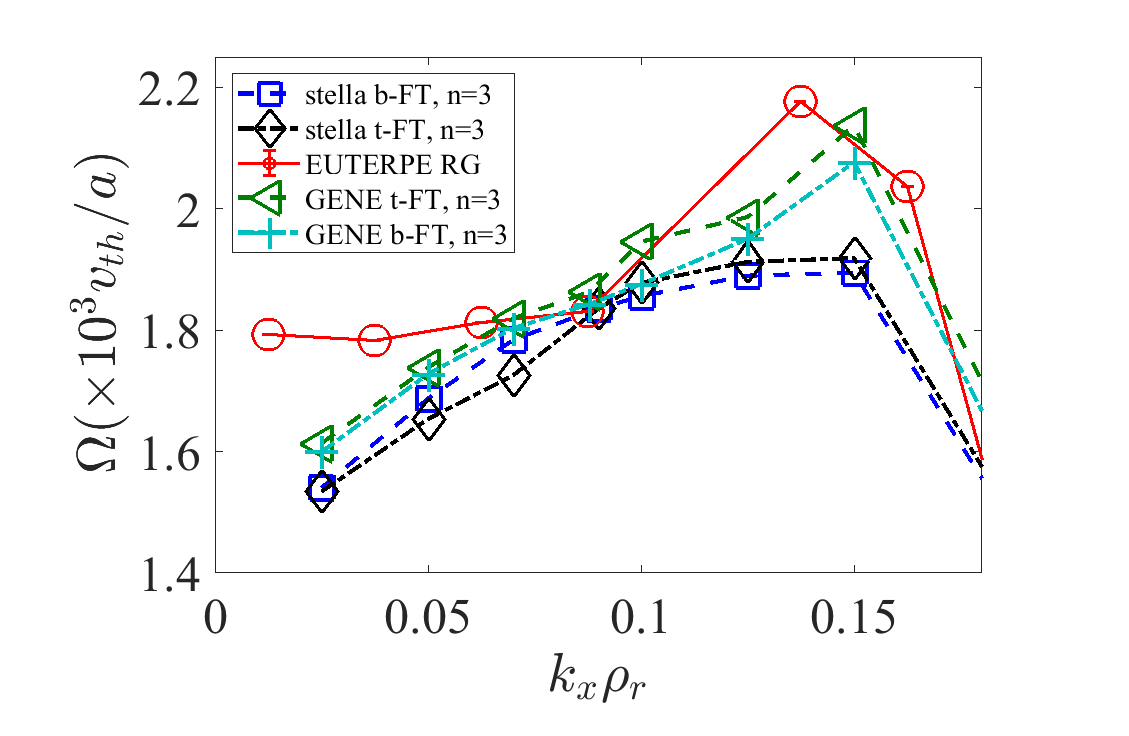}
	\caption{ZF oscillation frequency versus $k_x$ in the W7-X configuration at r/a=0.5 from simulations with GENE in FTs with lengths $n_{pol}=1-6$ (top). Comparison of ZF frequency in  FT simulations with GENE and \texttt{stella} with $n_{pol}=3$ and RG simulations with EUTERPE in W7-X (bottom).}
	\label{FigZFFreqnPotl3}
\end{figure}
One distinct feature that the ZF relaxation exhibits in stellarators is the low-frequency oscillation \cite{Mishchenko08}, which is related to the average radial drift of trapped particles \cite{Monreal17}. The characteristic ZF oscillation frequency is much smaller than that of the geodesic acoustic mode (GAM) \cite{Winsor68}  and increases with the averaged radial drift of trapped particles \cite{Monreal17}. 
\nuevo{Figure 
	\ref{FigZFFreqnPotl3} shows} the comparison of the oscillation frequency obtained from calculations in the FT and RG computational domains in W7-X. Note that in LHD the ZF relaxation is dominated by the GAM oscillations, which are weakly damped due to its small  rotational transform, and the low frequency oscillation is hardly appreciated.
In the case of W7-X the situation is {the} opposite, the GAM is strongly damped and the relaxation is dominated by the low-frequency oscillation. The low-frequency oscillation is {also} strongly damped as the radial wavenumber increases \cite{Helander11} and then it is only safely captured for small wavenumbers. 
\nuevo{The top panel in Figure \ref{FigZFFreqnPotl3} shows the ZF frequency vs $k_x$ for several FTs with increasing length from {$n_{pol}=1$ to $n_{pol}=6$}.} As for the residual level, the oscillation frequency obtained in FT simulations with $n_{pol}=1$ is significantly \nuevo{larger} than that obtained in a RG domain and the results approach the RG result as the length is increased up to {$n_{pol}=3,6$. Again, for $n_{pol}=4,5$ the oscillation frequency obtained in FT simulations separates from the RG result more than for $n_{pol}=3,6$, although these deviations are much smaller than in the case of the ZF residual}. 

\begin{figure}
	\centering
	\includegraphics[trim=30 66 40 0, clip, width=8.5cm]{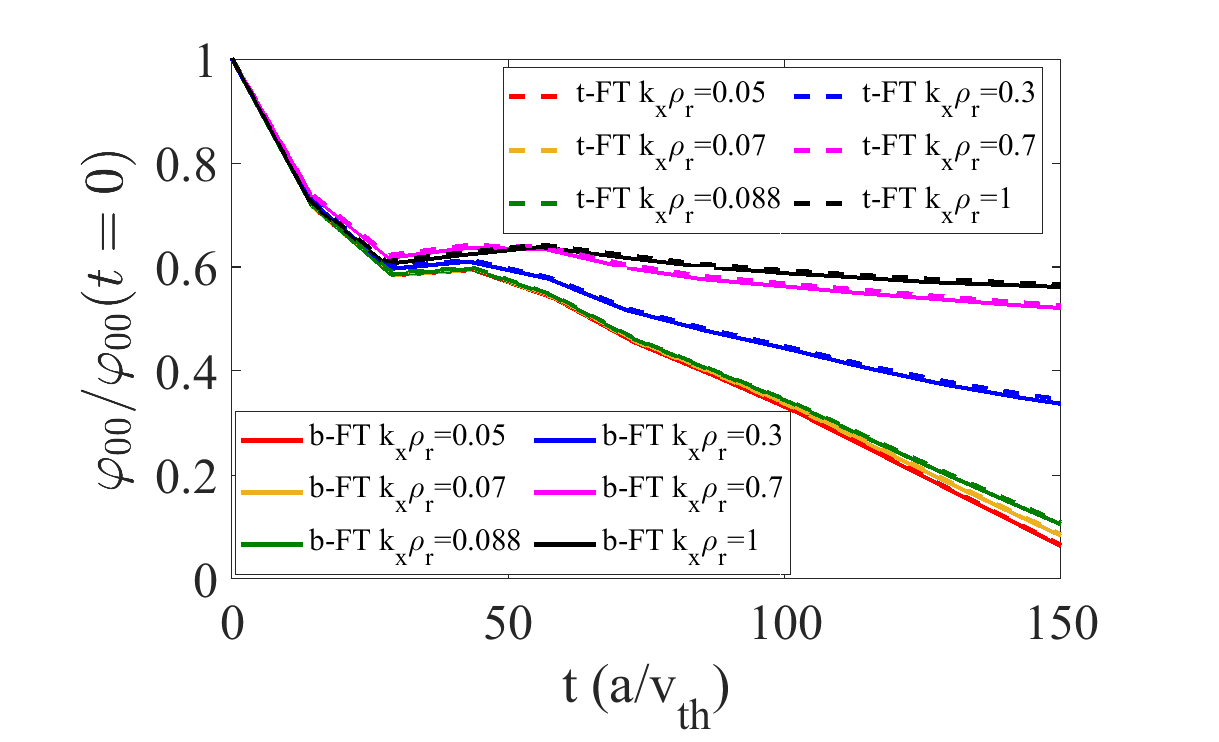}\\
	\includegraphics[trim=30 0 40 20, clip, width=8.5cm]{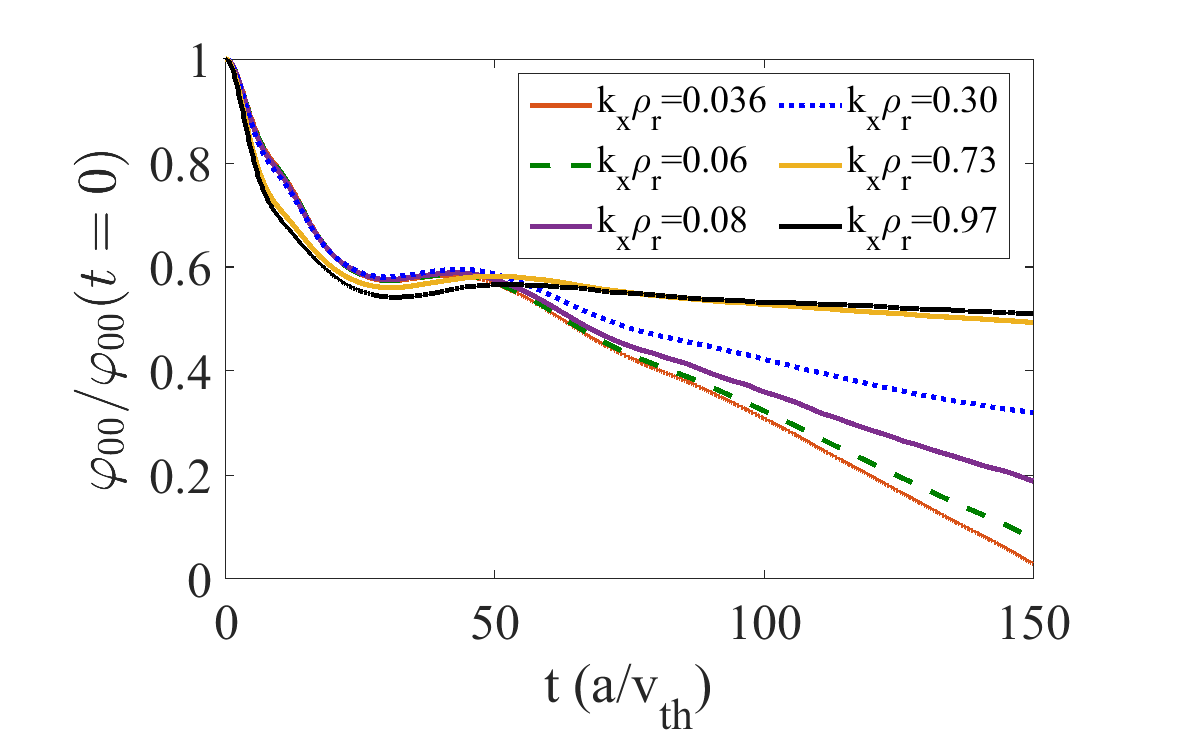}
	\caption{Comparison of the short-time evolution of potential in FZ relaxation simulations with GENE for the FT (top) and RG (bottom) computational domains in W7-X. All the FT simulations correspond to $n_{pol}=3$.}
	\label{FigZFShortTime}
\end{figure}

	\nuevo{The residual zonal flow level is a long-time-scale property and it is questionable that this value can affect the nonlinear saturation of turbulence,  which occurs in much shorter times. However, the short-time relaxation of the potential could more likely affect the saturation process. Here we look at the short-time scale behavior of the zonal  potential decay  in different computational domains. 
	Figure \ref{FigZFShortTime} shows the short-time evolution of the zonal potential for several radial scales in simulations of ZF relaxation in the W7-X configuration for the two FTs with $n_{pol}=3$ and that of the radial electric field\footnote{The electrostatic potential shows a similar, although noisier, time evolution than the electric field, which is the reason to use the later for fitting the model in Eq. \ref{EqModelFit} for extracting $R$ and $\Omega$.} for several RG simulations with EUTERPE for different radial scales of the initial perturbation. At very short times, below $\sim 50 ~a/v_{th}$ the decay is almost independent of the radial scale of the perturbation  in both computational domains, but for longer times, $t>50 ~a/v_{th}$, it  is clear that the short-time evolution changes with the radial scale as the residual level does. The similarity in the decay between both domains is remarkable.
	}
	\nuevo{No significant differences are found in the short time decay of potential between FTs with different lengths in \texttt{stella}, while in GENE FT simulations  a difference  between FTs (not shown) is found for small values of the radial scale $k_x$.}

\nuevo{In conclusion, it} is clear that short flux tubes are not {sufficient} to capture the long-term properties of the ZF relaxation, in general. 
Different flux tubes provide different results and they converge to {the same} result when the length of the FT is increased. The number of poloidal turns in the flux tubes that is required for convergence depends on the magnetic geometry and a general rule {applicable to} other magnetic configurations cannot be extracted from present practical cases. \nuevo{In the short-time evolution of the zonal potential less clear difference between flux tubes is found than in the long-term properties.}

			\section{Linear stability of ITG modes}\label{SecLinearITGS}
 Now we turn {our view to the} ITG instability in different computational domains: FT, FFS and RG. We run simulations with adiabatic electrons and  follow a similar approach to that {employed} in the previous section. Simulations for two stellarator-symmetric FTs {with different lengths} are compared with each other and also with FFS and RG simulations. \nuevo{We use the same codes,  which provide} different levels of information about the mode structure. Global codes run simulations with multiple modes whose radial structure can 
 be extracted by Fourier transforming the radial mode structure in {the} post-processing.
Full{-flux-}surface radially local simulations are multi-mode in the perpendicular direction along the flux surface, thus considering multiple $k_y$ scales. In the radial direction, GENE-3D contain also a number of different radial scales $k_x$, determined by the box size. The radial structure of the modes could also, in principle, be extracted. In GENE, one single $k_x$ mode can be simulated.
On the other hand, local linear FT simulations can scan the stability for each $k_y$  mode separately. With respect to the radial scale of the modes, in FT simulations, $k_x$ is set as a parameter and a set of simulations with different values of $k_x$ \nuevo{and $k_y$} has to be run in order to study {how the linear instability depends on the radial scale}. 

The basic quantities to compare here will be the growth rate $\gamma$ and the frequency $\omega$ of the ITG modes, which can be extracted from all the codes. The local FT codes can provide the growth rate and frequency for each $(k_x, k_y)$ pair separately, while from the RG and FFS simulations\nuevo{, eventhough they consider multiple modes,} only the  frequency and the linear growth rate of the most unstable mode considered in the simulation, which determines the growth of both mode and numerical noise amplitudes, can be extracted. Using the phase factor in EUTERPE, an estimate of the growth rates and frequencies of different modes can be scanned in a set of simulations in which the center of the Fourier filter is varied.       
For the comparison of wavenumbers the relation $k_y=\frac{m}{r}$, with $m$ the poloidal mode number in PEST coordinates and $r$ the {effective minor} radius, is used. The $k_y$ values are normalized with the {inverse} Larmor radius, as usual. The growth rates and real frequencies are normalized to $v_{th}/a$, with $v_{th}=\sqrt{2T/m}$ the ion thermal velocity.

\begin{figure}
	\centering
	\includegraphics[trim=5 75 0 20, clip, width=8.25cm]{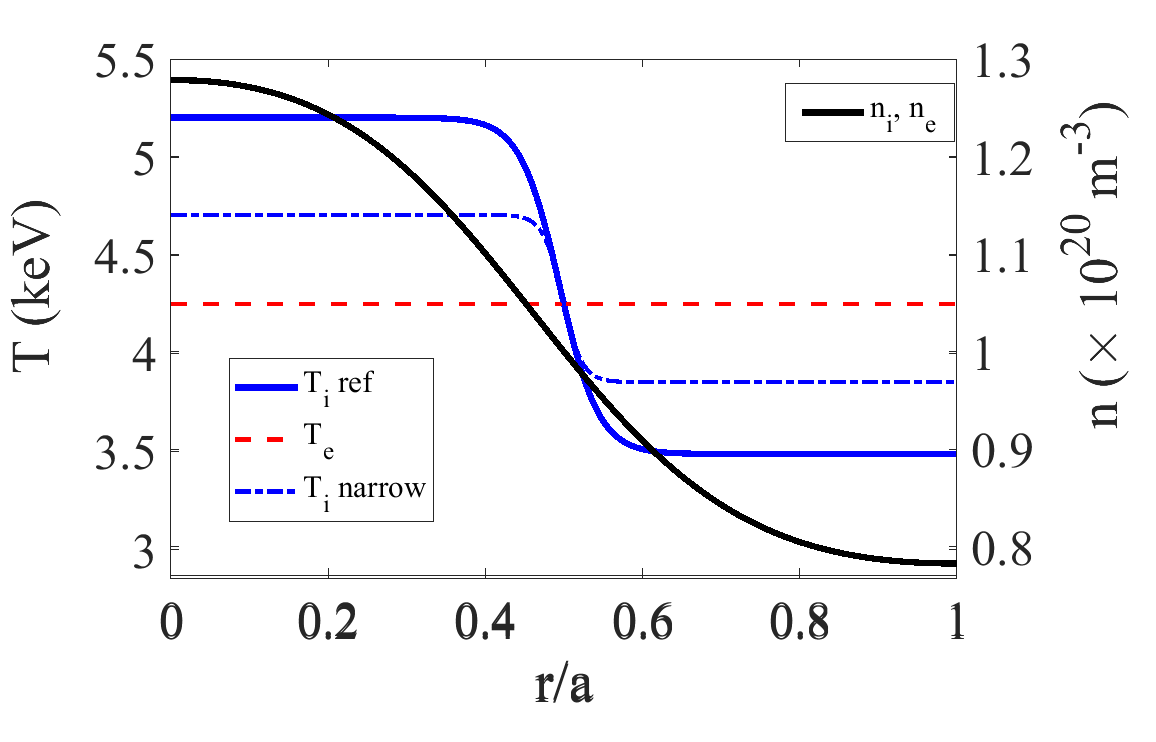}\\
	\includegraphics[trim=5 5 0 20, clip, width=8.25cm]{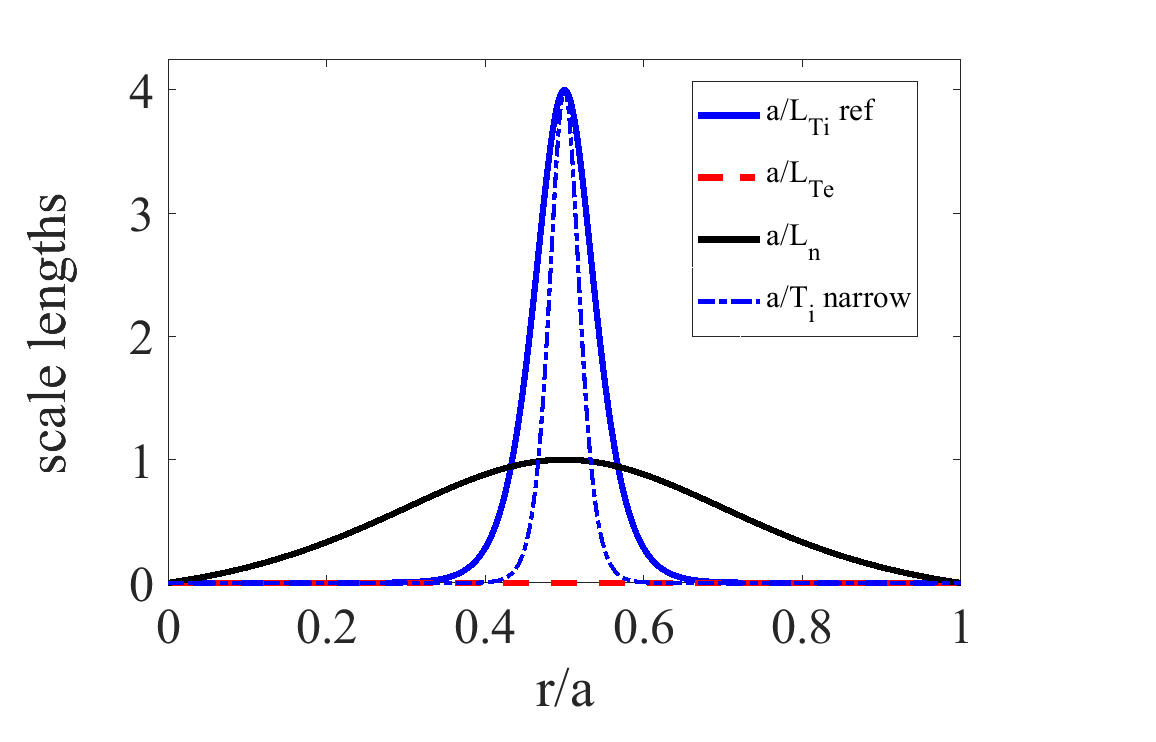}
	\caption{Density and temperature profiles used in simulations of ITGs (top) both in LHD and W7-X configurations and scale lengths of density and temperatures (bottom).}
	\label{FigProfiles}
\end{figure}
\nuevo{
The settings of the simulations are as follows. The electron temperature profile is flat, with $T_e=4.25 ~\rm{keV}$, and density and ion temperature profiles, similar to those in \cite{Sanchez20} are  used, which are given by the formula
\begin{eqnarray}
X = X_* \rm{exp} \left[\frac{-\kappa_x}{1-\rm{sech}^2_{x} } \left( \tanh(\frac{r - r_0}{a\Delta_x}) - \rm{sech}^2_{X} \right)\right],
%
\label{analProfilesnT}
\end{eqnarray}
where $X=\{n,T_i\}$ represents a radial profile of density ($n_e=n_i=n$) or ion temperature, $n_*=10^{20}~m^{-3}$, $T_*=4.25~\rm{keV}$, $r_0=0.5a$, $\kappa_{Ti}=4$, $\kappa_n=1$, $\Delta_n=0.3$ and $\rm{sech}^2_{X}=\cosh^{-2}(\frac{r_0}{a\Delta_X})$ and $a$ the minor radius. The profiles used in these simulations are shown in Figure ~\ref{FigProfiles}. 
Two profiles of $T_i$ are shown; first, the one used in all the ITG simulations presented in Figures \ref{FigGRFR1} and \ref{FigGRFR2}, with $\Delta_{Ti}= 0.1$, shown with continuous thick line and labeled as ``ref" in the figure; 
and also a profile, labeled as ``narrow", with $\Delta_{Ti}=0.08$,  which has a smaller  width of the $\eta_i$ profile.
The local values at $r/a=0.5$, $T_i(r/a=0.5)=T_e=T_*$, $L_n(r/a=0.5)=1$ and $L_{Ti}(r/a=0.5)=4$,  are used as input for the radially local simulations (FT and FS), all carried out at this radial position.  
}

With these settings, linear simulations with adiabatic electrons are carried out in the standard configuration of LHD in the FT, FS and RG domains and the growth rate $\gamma$ and the real frequency $\omega$ obtained in different domains with the different codes are compared. \nuevo{The numerical settings used in all the simulations in this section are given in Table \ref{tabITGs}.}

\begin{figure}
	\centering
	\includegraphics[trim=20 65 60 20, clip, width=8.25cm]{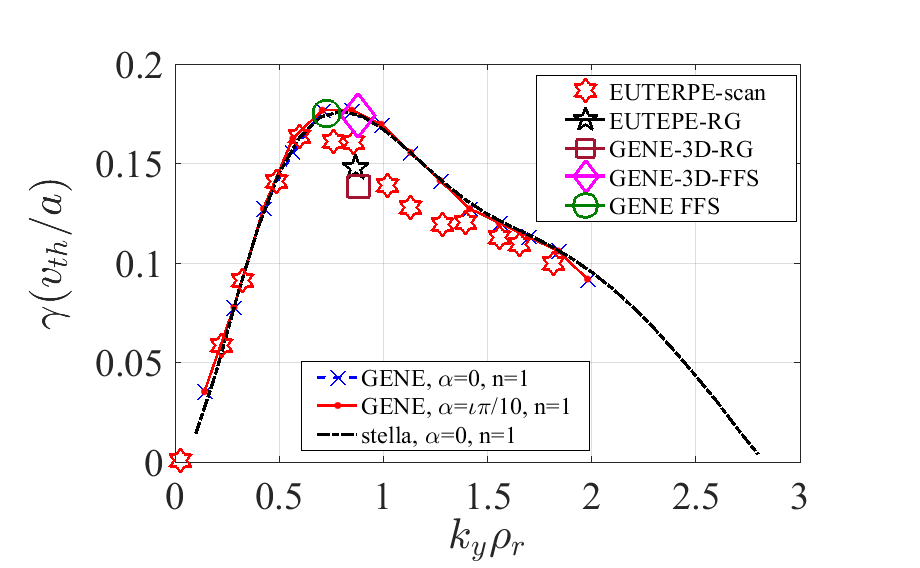}
	\includegraphics[trim=20 5 60 20, clip, width=8.25cm]{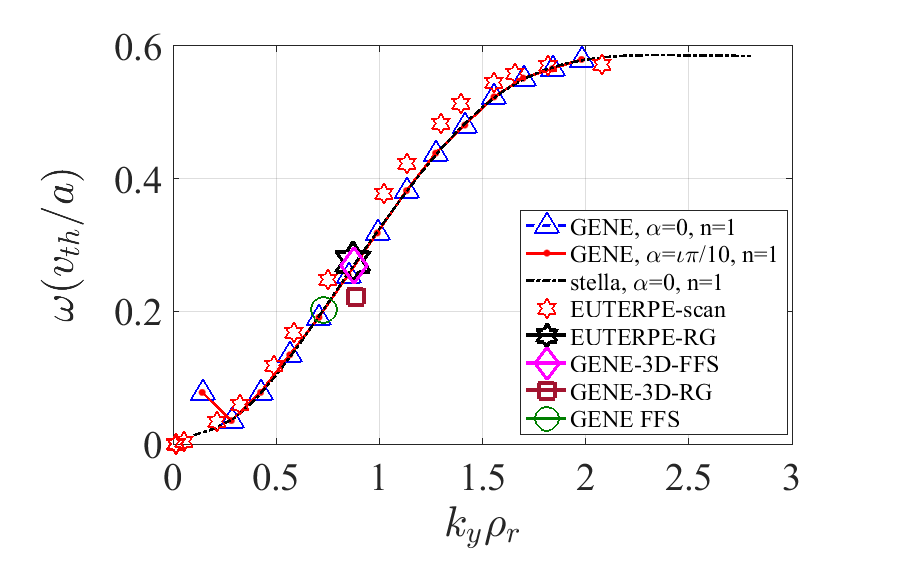}
	\caption{Comparison of growth rate $\gamma$ (top) and frequency $\omega$ (bottom) vs the normalized wavenumber $k_y\rho_r$ for FT, FFS and RG simulations in the standard configuration of LHD at $r/a=0.5$.}
	\label{FigGRFR1}
\end{figure}

 Figure \ref{FigGRFR1} shows the growth rate and frequency obtained in different computational domains in the LHD configuration at $r/a = 0.5$. \nuevo{FT simulations were carried out with both GENE and \texttt{stella} and the agreement between both codes is excellent. Only simulations in FTs with $n_{pol}=1$ are shown in the Figure for clearness, but FTs with lengths $n_{pol}=2,3$ provide almost exactly the same results for both  $\gamma$ and $\omega$}. FFS simulation results obtained with \nuevo{GENE\footnote{The FFS simulation with GENE has been carried out with $T_*=1.5 ~\rm{keV}$ because of technical reasons.} and} GENE-3D,  and results from RG simulations with EUTERPE are also shown in this figure.  In FFS  and RG simulations only one point corresponding to the most unstable mode is represented. \nuevo{In the FFS simulation with GENE only one $k_x$ is resolved, while in GENE-3D a set of $k_x$ are resolved}. The \nuevo{wider resolution simulation with EUTERPE (labeled as ``WR" in the table)} {is full volume and} covers the wavenumbers $1<m<127$, 
which allows resolving modes up to $k_y\rho_r< 1.58$.  In addition to the full {volume} simulation, a scan in wavenumber has been carried out with EUTERPE. In this scan, the radial domain is limited to $0.31<r/a<0.7$ and the Fourier filter keeps modes with $m_0 -31 < m < m_0 + 31$, with $m_0$ {varied} to scan different regions of the wavenumber spectrum. A FFS simulation was carried out with GENE-3D
 at r/a=0.5, with radially periodic boundary conditions, using the following resolution: $(n_x, n_y, n_z, n_{v\parallel}, n_{\mu})=(90, 64, 128, 32, 8)$ and resolving modes $k_y\rho_r<9$. Here, $(n_x, n_y, n_z, n_{v\parallel}, n_{\mu})$ are the number of points in the grids along the $x$, $y$, $z$, $v_{\parallel}$ and $\mu$ coordinates. For the RG simulation with GENE-3D, the radial domain $0.2<r/a<0.8$ was covered, the resolution was  $(n_x, n_y, n_z, n_{v\parallel}, n_{\mu})= (144,64,128,48,20)$ and covered the range $k_y\rho_r<9$. 
 A very good agreement between all FTs and FFS domains is obtained, which can be explained by the smoothness and approximate periodicity along the magnetic field lines of the magnetic quantities in LHD. \nuevo{Figure \ref{FigMagFielQntts} shows the magnetic field strength along a field line followed by the $\alpha=0$ FT in the LHD configuration}. The 
 RG simulations provide a $\gamma$ for the most unstable mode that is smaller than that obtained in FT or FFS simulations by 15-20\%. The agreement  in the mode frequency between all the codes/domains is very good.

\begin{table*}[h]
			\centering		
			\footnotesize 
				\caption{\label{tabITGs}Numerical settings used in the ITG simulations in this paper. $^\dagger$$T_*=1.5 ~\rm{keV}$ was used for this simulation due to technical reasons. $^*$The resolutions in the MHD equilibrium magnetic field and the perturbed electrostatic field are independent in EUTERPE, while in \texttt{stella}, GENE and GENE-3D they are coupled. $^{**}$In EUTERPE, $n_x, n_y,n_z$ are the resolutions in the $s$, $\theta$, and $\phi$ PEST coordinates, respectively.}
				\noindent
				\centering		
			\begin{tabular}[h]{cccccccccccc}
				\hline	
				&  \multicolumn{2}{c}{\hspace{.2cm} code \hspace{.2cm}  domain} &  sim.  & $\Delta m$ & $r/a$ & $n_x^{**}$ & $n_y^{**}$  & $n_z^{**}$ & $n_{v_{\par}}$ & $n_{\mu}$ \cr 
				\hline  
				\hline
				\multirow{6}{*}{{LHD}}	 & \texttt{stella}  & {FT}  &  & & 0.5 &   1   & 1  & 128 & {48}  & {24} \cr 
				\cline{2-12}
				& \multirow{2}{*}{GENE}  & {FT}   &  &  & 0.5 & 1 & 1 & $256\times n_{pol}$ &  48& 12\cr 
				\cline{4-12}
				&  						 & {FFS{$^\dagger$}}   &  &  & 0.5 &12 & 128 & 256 &  32& 12\cr 
				\cline{2-12}
				& \multirow{2}{*}{GENE-3D} & FFS    & &  & 0.5 & 90 & 64 & 256 & 16 & 8 \cr 
				\cline{4-12} 
				&  & RG  & &  & 0.2-0.8 & 144 & 64 & 128 &  48 & 20 \cr  
				\cline{2-12}
				& \multirow{3}{*}{EUTERPE} & \multirow{3}{*}{RG$^*$}	& ${\varphi}$ LR 	& 127 & 0-1& 64	& 256 & 64  \cr 
				\cline{4-12}
				&	&	& ${\varphi}$ scan &  63 	& 0.31-0.7 & 96	& 64 & 32  \cr  
				\cline{4-12}
				&  & & MHD  &  &  0-1& 141	&  256 & 256 &  \cr
				\hline
				\hline\cr
					\hline
					\hline				
				\multirow{12}{*}{{W7-X}}	& \multicolumn{1}{c}{\texttt{stella}} & {FT}   &  &  & 0.5 &   1   & 1  & $128\times n_{pol}$ & 48  & 24\cr
				\cline{2-12} 
				& \multirow{3}{*}{GENE}  & {FT}   &  &  & 0.5 & 1 & 1 & $128\times n_{pol}$ &  48 & 12 \cr
				\cline{4-12} 
				&   & { FFS} &    &  &  \multirow{3}{*}{0.5} &  12	&  256	& 128	& 32 & 12\cr
				\cline{7-12} 
				&   & { FFS} &    &  &   &  12	&  96	& 128	& 32 & 12\cr
				\cline{2-12} 
				& \multirow{3}{*}{GENE-3D} & \multirow{3}{*}{ FFS}  & &   & \multirow{3}{*}{\nuevo{0.5}} & 90 &  384 & 128 & 32 & 8&   \cr
				\cline{7-12} 
				&  &   & &   &   & 90 &  192 & 128 & 32 & 8&   \cr
				\cline{7-12} 
				&  &   & &    &    & 90 &  96 & 128 & 32 & 8&  \cr
				\cline{2-12} 
				& \multirow{3}{*}{EUTERPE} &  \multirow{4}{*}{RG$^*$} & ${\varphi}$ WR   & 1023	& 0.31-0.7& 256	& 1024 & 256  \cr
				&	& & ${\varphi}$  MR  & 255	& 0-1 & 128	& 256 & 64  \cr
				\cline{4-12} 
				&	& & ${\varphi}$ scan &  34	& 0.31-0.7 & 128	& 64 & 64  \cr
				\cline{4-12} 
				& & &  MHD &   & 0-1& 99	& 256 & 256 &  \cr
				\hline
				\hline
			\end{tabular}
	
\end{table*}
\begin{figure}
	\centering
	\includegraphics[trim=30 80 60 10, clip, width=8.25cm]{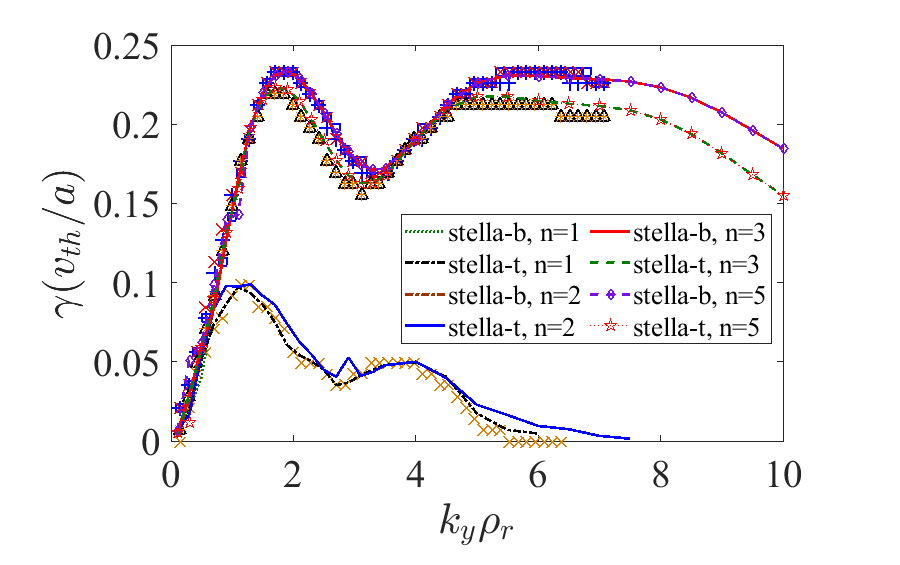}
	\includegraphics[trim=30 10 60 5, clip, width=8.25cm]{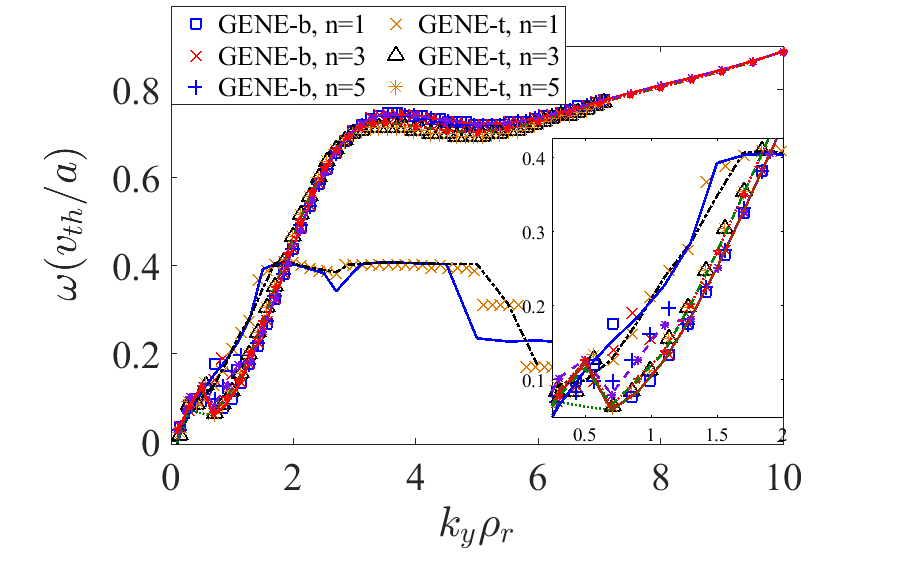}
	\caption{Comparison of growth rate $\gamma$ (top) and frequency $\omega$ (bottom) vs the normalized wavenumber $k_y\rho_r$ for FT simulations with GENE and \texttt{stella} in the KJM configuration of W7-X with beta=3\% at $r/a=0.5$. The simulations for the bean and triangular flux tubes are labeled as ``-b" and ``-t", respectively. The inset in the bottom panel shows the frequencies for the range $0<k_y\rho_r<2$.}
	\label{FigGRFW7X-FTs}
\end{figure}
Figures \ref{FigGRFW7X-FTs} and  \ref{FigGRFR2}  show the comparison of $\gamma$ and $\omega$ vs. the normalized  wave number for FT, FFS and RG simulations in the KJM configuration of W7-X. Simulations for the two stellarator-symmetric FTs,  $\alpha=0$ (bean) and $\alpha=\iota\pi/5$ (triangular) \nuevo{with lengths $n_{pol}=1-5$, carried out with GENE and \texttt{stella} are shown in Figure \ref{FigGRFW7X-FTs}}. 
  It is clear from the figure that in W7-X the situation is quite different from that for LHD, shown in Figure \ref{FigGRFR1}. The different FTs, with the shortest lengths ($n_{pol}=1,2$), provide different results for both $\gamma$ and $\omega$. \nuevo{The disagreement between short FTs is found both for the growth rate and the frequency.  In the range $0<k_y\rho_r<2$ the growth rates in different FTs are very similar but there is a clear disagreement in the frequencies, as highlighted in the inset of the bottom panel of Figure \ref{FigGRFW7X-FTs}. In this region, a clear difference is also found in the radial scale of the most unstable modes for the bean and triangular FTs (see section \ref{subSecDepRadScale}). The difference in the results for short FTs} can be traced back to the fact that the short bean and triangular FTs explore regions of the flux surface with very different properties \nuevo{(see Figures \ref{FigMagFielQnttsOnePeriodNpol1} and \ref{FigMagFielQnttsOnePeriod})}. As the length is increased, the FTs explore a larger region of the flux surface and the results from both FTs get closer, although bean and triangular FTs results do not match exactly even for large values of $n_{pol}$. Cases with $n_{pol}=5$ and $10$ (not shown) demonstrate that good convergence with the length for each flux tube is found for $n_{pol}=3$ with both \texttt{stella} and GENE, however, both FTs do not match with each other. 
  \nuevo{Figure \ref{FigGRFR2} shows the comparison of $\gamma$ and $\omega$ in the W7-X configuration for the FT ($n_{pol}=3$), FFS and RG domains.}  
A wavenumber scan with EUTERPE using \nuevo{the density and temperature reference profiles shown in Figure \ref{FigProfiles}} and a small Fourier filter ($m_0 -31 < m < m_0 + 31$) is also shown in this figure, which qualitatively reproduces the wavenumber dependency from FT results, although with growth rates significantly smaller than those {obtained for the bean FT in the region} $1<k_y\rho_r < 3$. In this scan, the radial domain is limited to $0.31<r/a<0.7$. \nuevo{The scan in wavenumbers has been repeated, using the same numerical settings and the 
	narrow profile from Figure \ref{FigProfiles} to check the influence of the width of the $\eta_i$-profile, if any. 
	For the narrow profile, only a slight reduction in $\gamma$ is found while the general trend is kept}. Finally, RG and FFS simulation results are shown with just a point per simulation, corresponding to the most unstable mode. \nuevo{Two more RG simulations with EUTERPE using {the reference profiles and} a wider resolution are shown. The simulation with the largest resolution (labeled as ``WR" in the table) uses a Fourier filter keeping modes with $1<m<1023$, 
corresponding to $0<k_y\rho_r<15$, and cover the full radius. The medium resolution one (``MR"  in the table) uses a filter keeping modes $1<m<255$ and $0<k_y\rho_r<3.75$.} 
Three FFS simulations with different resolutions are shown for GENE-3D. The largest-resolution case, with resolutions  $(n_x, n_y, n_z, n_{v\parallel}, n_{\mu})=(90, 384, 128, 32, 8)$, covers the wavenumbers $0<k_y\rho_r<17$, and for the others, the resolution $n_y$ is reduced by factors 2 and 4 and the largest wavenumbers resolved are $8.5$ and $4.25$, respectively. Similar to what happens with EUTERPE simulations, the most unstable mode is found at wavenumbers increasing with  resolution. \nuevo{Two FFS simulations carried out with GENE  using different resolutions and resolving the maximum wavenumbers  $k_y\rho_r<3.9$ and 10.5, respectively, are also shown in the same figure. The relevant numerical settings used in all these simulations are given in Table \ref{tabITGs}. The results of GENE FFS simulations reasonably agree with FT simulations with GENE and \texttt{stella} showing a slightly larger growth rate. In GENE FFS simulations only the radial wavenumber $k_x=0$ is considered, while in GENE-3D a radial grid is used, which allows resolving a set of finite scales $k_x$.}

\begin{figure}
	\centering
	\includegraphics[trim=30 80 60 20, clip, width=8.25cm]{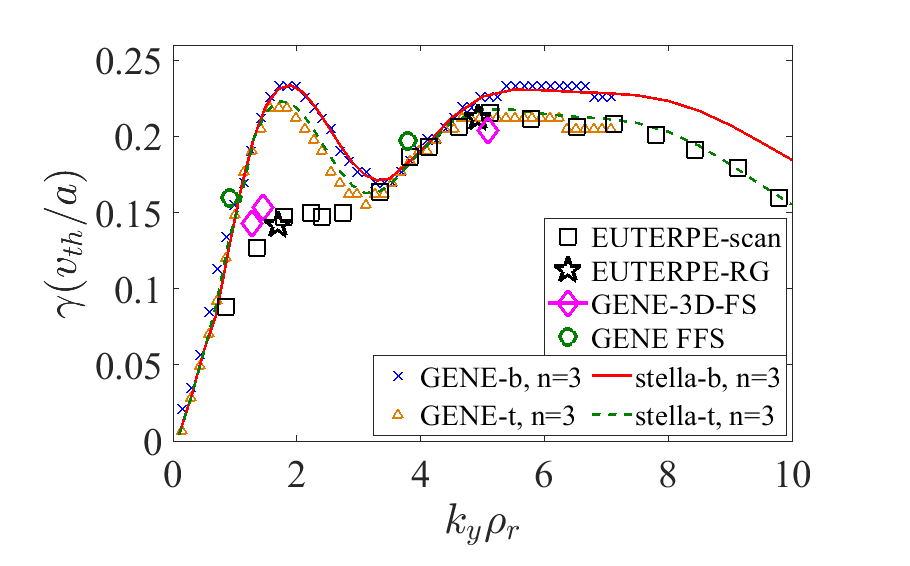}
	\includegraphics[trim=30 10 60 20, clip, width=8.25cm]{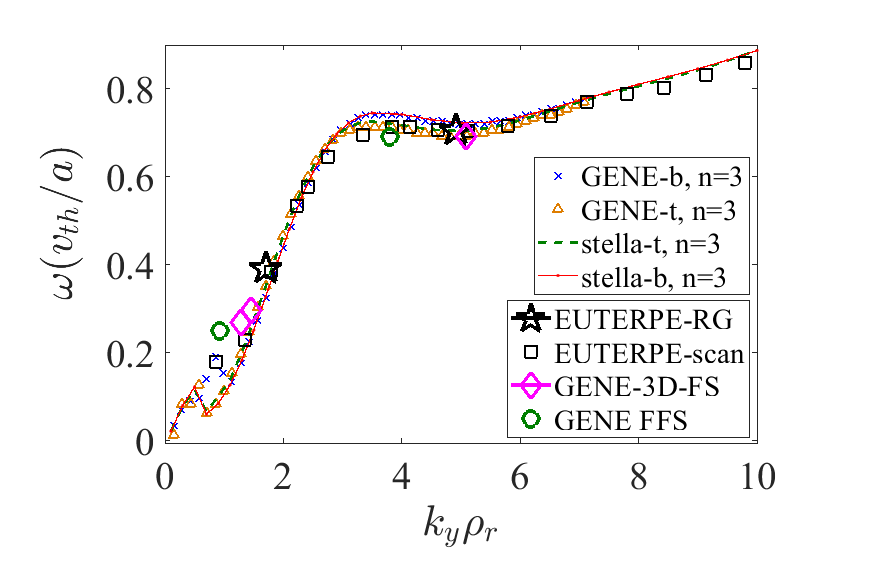}
	\caption{Comparison of growth rate $\gamma$ (top) and frequency $\omega$ (bottom) vs normalized wavenumber $k_y\rho_r$ for FT, FFS and RG simulations in the KJM configuration of W7-X with beta=3\% at $r/a=0.5$.}
	\label{FigGRFR2}
\end{figure}
 For $k_y\rho_r>3$, the results for FFS with GENE-3D and RG with EUTERPE reasonably coincide in the wavenumber of the most unstable mode, with $\gamma$ close to that of \nuevo{ the triangular FT} and slightly smaller for FFS than for RG. In the range $k_y\rho_r<3$, both FFS and RG simulations give values of $\gamma$ in between the most (triangular) and less unstable (bean) flux tubes. \nuevo{ Interestingly, the peak in the growth rate found around $k_y\rho_r \sim 2$ by FT simulations is neither captured by the FFS nor the RG simulations.}

For the frequencies, the {agreement} between all codes is very good for wavenumbers $k_y\rho_r>2$ and for FTs with $n_{pol}>3$ in both \texttt{stella} and GENE. In the range $k_y\rho_r<2$, the agreement between domains \nuevo{and between different FTs} is not so good (see also Figure \ref{FigGRFW7X-FTs} for the detailed comparison of FTs). 

The lack of periodicity of magnetic quantities along the field line  in \nuevo{the W7-X configuration (see Figure  \ref{FigMagFielQntts})} makes that different FTs give different results, in general. When the FT length is increased, the triangular flux tube results get close to those of the bean one. This is explained by the fact that \nuevo{for $n_{pol}=3$ the triangular FT gets closer to the regions explored by the bean FT with $n_{pol}=1$, which finds the most unstable region around $(\theta,\zeta)=(0,0)$  (see Figure \ref{FigMagFielQnttsOnePeriodNpol1}, \ref{FigMagFielQnttsOnePeriod} and \ref{FigMagFielQntts})}.    
A scan of FT simulations in the radial wavenumber \nuevo{(see section \ref{subSecDepRadScale}) shows that for the triangular FT, the most unstable mode is not always that of $k_x=0$ {and that the scale $k_x$ of the modes at which the maximum growth rates are obtained changes with the FT length.} }

		%
		\subsection{Dependence with the radial scale}\label{subSecDepRadScale}
		%
		\nuevo{In this section we study the radial scale of unstable ITG modes in FT simulations carried out with \texttt{stella} and compare them with RG simulations with EUTERPE. As described in section \ref{secCodesAndDomains},  in linear FT simulations $k_x$ is a parameter while in RG in RG, since the radial is not not a spectral coordinate, a continuum of radial scales is simulated.many radial scales are simulated at the same time. Then, in order to extract the dependence with $k_x$ for $\gamma$ and $\omega$ in a FT, a scan in $k_y$ and $k_x$ has to be done. From a linear RG simulation, the radial scale for the most unstable mode is extracted by using a Fourier transform. We have computed the radial scale $k_x$ for the wavenumber scan shown in Figure \ref{FigGRFR1} for LHD and a value $k_x\rho_r \sim 0.22$ is obtained for a wide range of wavenumbers $0<k_y\rho_r<2.5$. \nuevo{FT simulations with \texttt{stella} do not show a strong dependence of $k_x$ with $k_y$ either}. In the W7-X configuration the situation is different, and a radial scale dependence on $k_y$ is found.}
\begin{figure*}
	\centering
	\includegraphics[trim=25 70 0 90, clip, width=7.75cm]{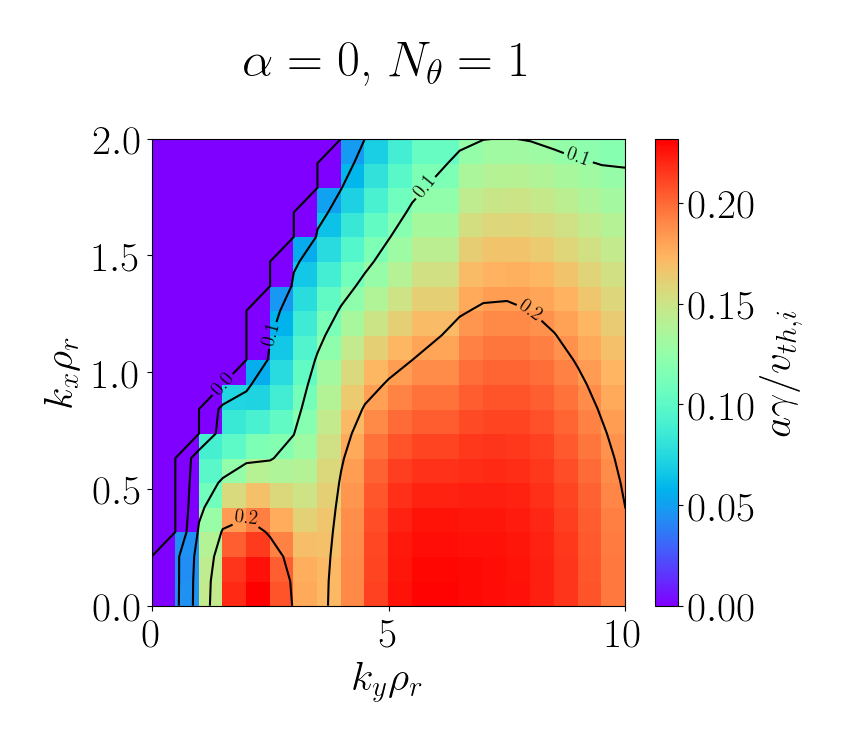}
	\includegraphics[trim=25 70 0 90, clip, width=7.75cm]{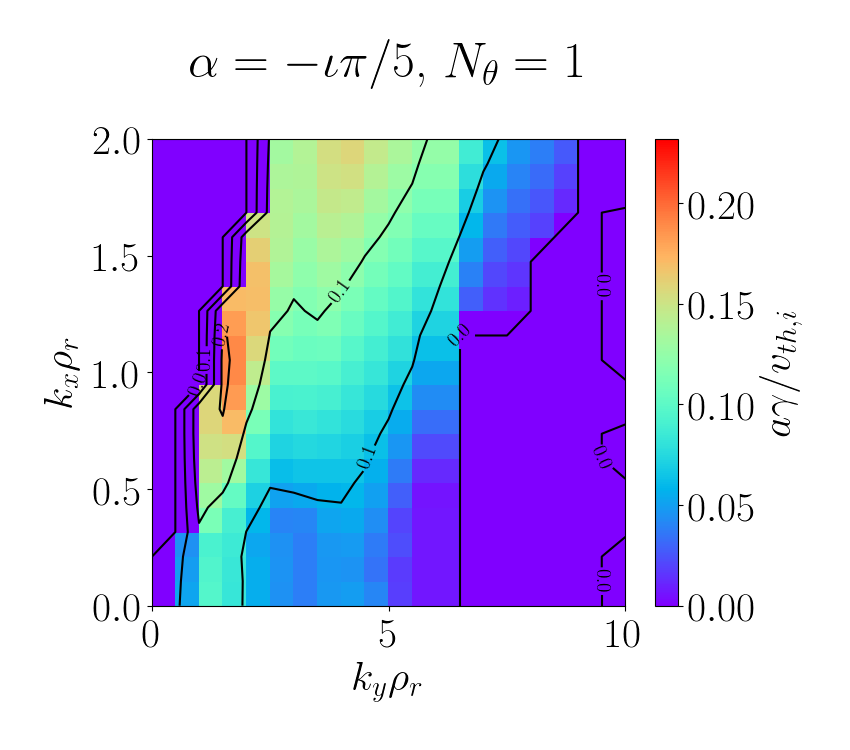}\\
	\includegraphics[trim=25 70 0 90, clip, width=7.75cm]{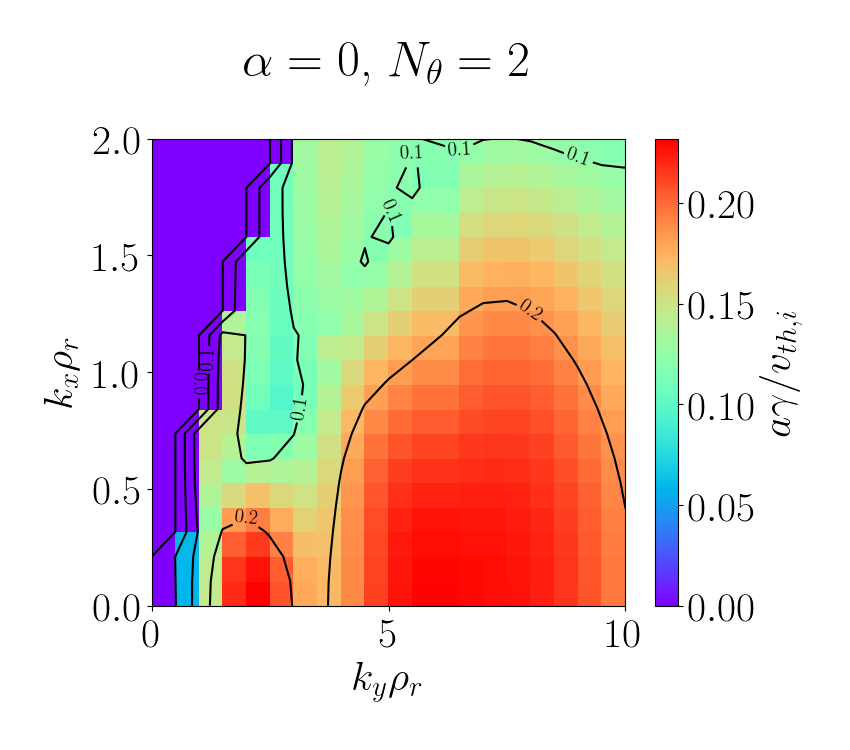}
	\includegraphics[trim=25 70 0 90, clip, width=7.75cm]{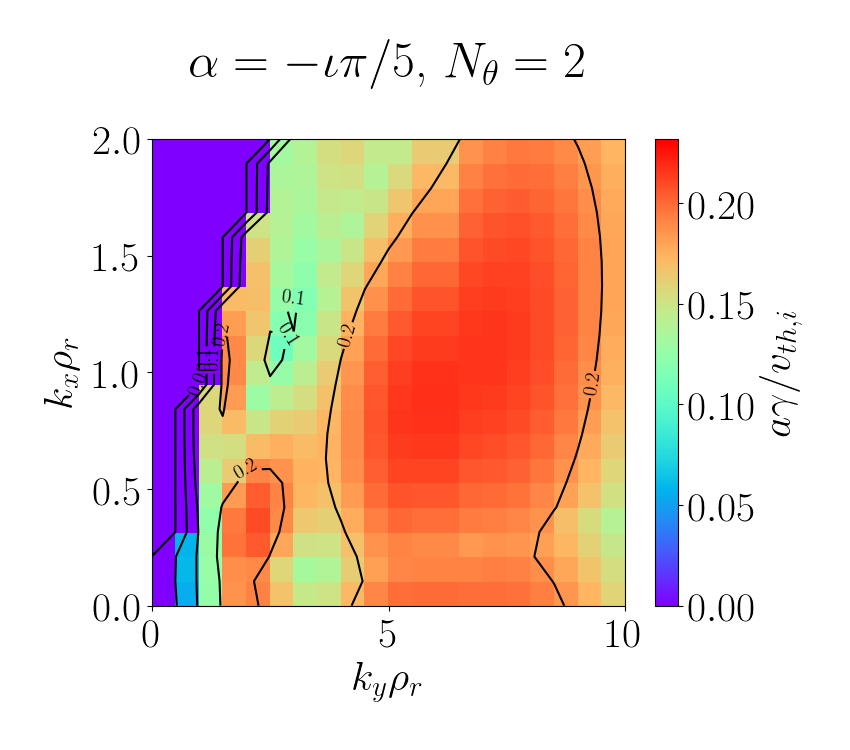}\\
	\includegraphics[trim=25 30 0 90, clip, width=7.75cm]{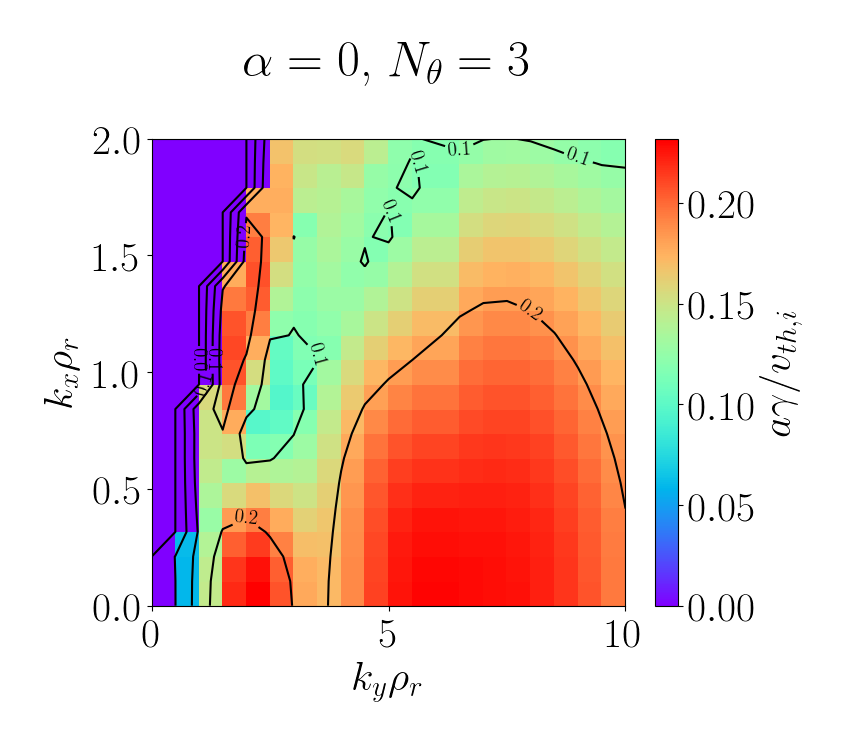}
	\includegraphics[trim=25 30 0 90, clip, width=7.75cm]{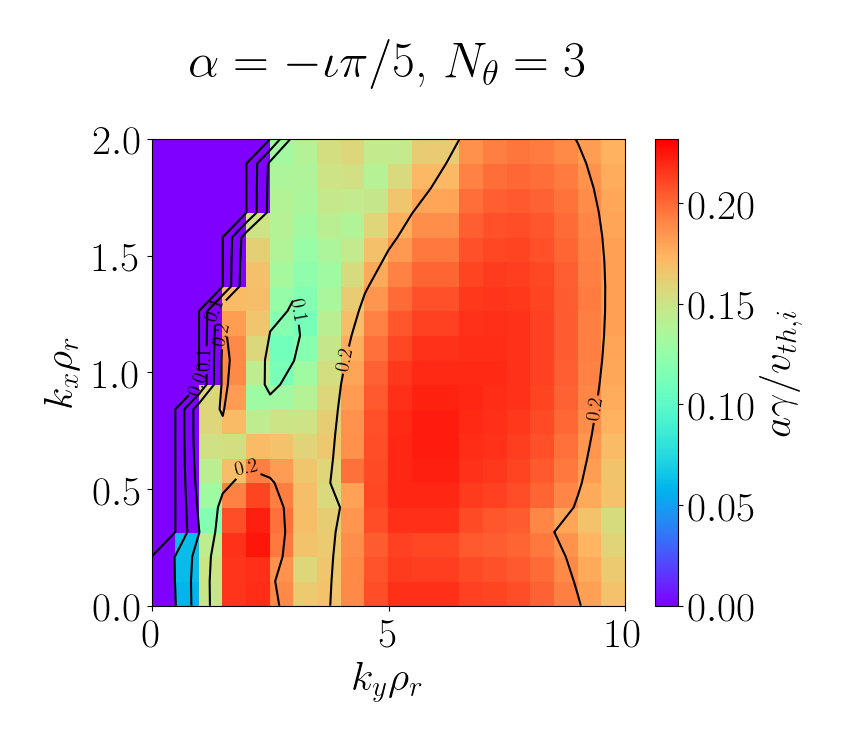}\\
	\caption{Growth rate vs. scales $k_x$ and $k_y$ in a scan carried out with \texttt{stella} for the KJM configuration of W7-X with beta=3\% at $r/a=0.5$. The left panels show the results for the bean FT and the right panels for the triangular one. Results for different lengths $n_{pol}=1$ (top),  $n_{pol}=2$ (middle) and  $n_{pol}=3$ (bottom) are shown. }
	\label{FigKxvsKyMapsStella}
\end{figure*}
	
\nuevo{Figure \ref{FigKxvsKyMapsStella} shows the growth rate in a scan in $k_x$ and $k_y$ carried out with \texttt{stella} in bean and triangular FTs with several  lengths, for the KJM configuration of W7-X. For short flux tubes, different results are obtained for the bean and triangular one. 
	The most unstable modes are located at $k_x \rho_r\sim 0$  in the bean flux tube, which is not the case for the triangular, where, unless the flux tube is extended so that it nearly overlaps with the bean flux tube, the most unstable mode is located at $k_x \rho_r\sim 1.5$ and $k_y\rho_r \sim 2$.
	When the length is increased to $n_{pol}=3$ these modes with  $k_x \rho_r\sim 1.5$ and $k_y\rho_r \sim 2$ appear also in the bean FT with a large growth rate, even more clearly than in the triangular FT at any length. 
}

	\begin{figure}
		\centering
		\includegraphics[trim=25 15 80 30, clip, width=7.5cm]{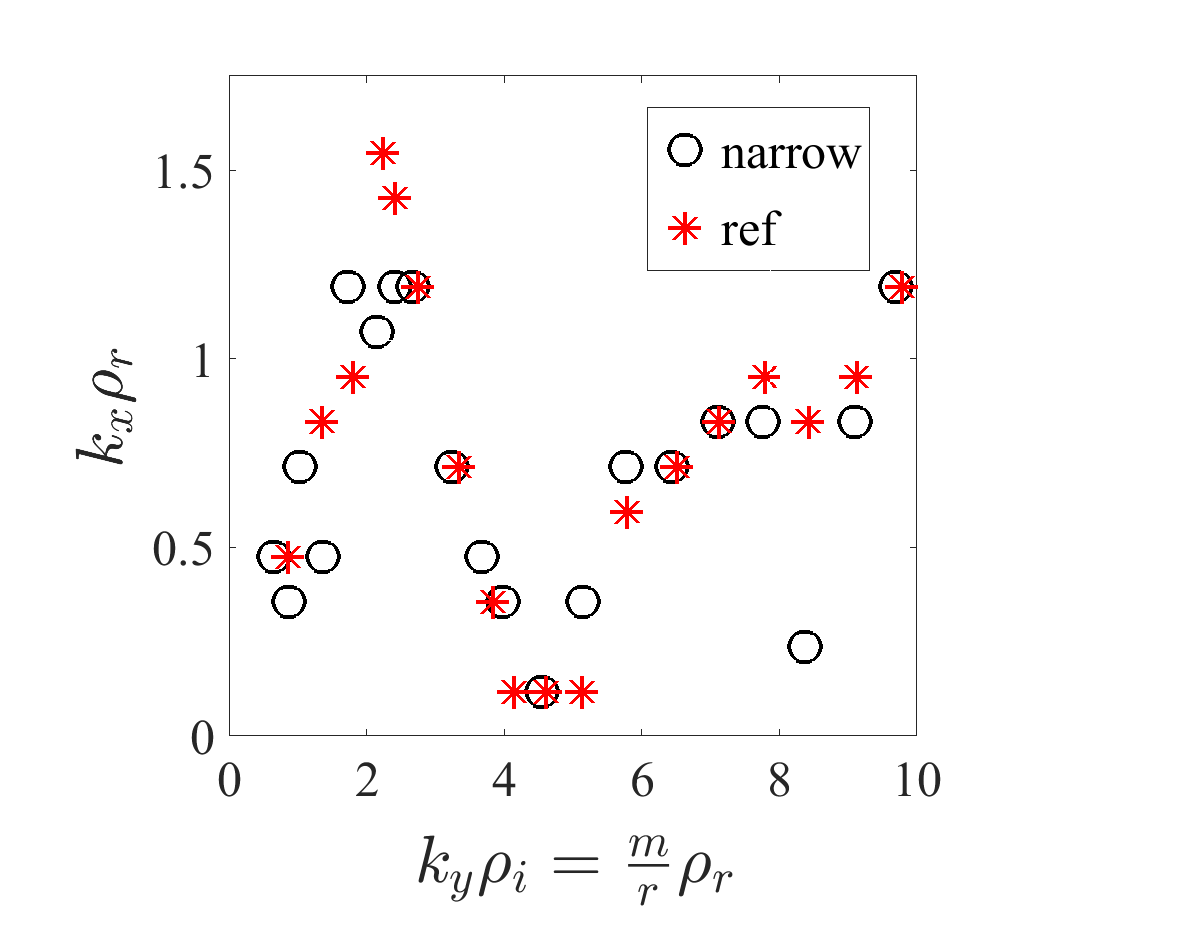}
		\caption{Radial scale $k_x$ of the most unstable mode in RG simulations with EUTERPE scanning for different $k_y$.}
		\label{FigKxvsKy}
	\end{figure}
				\nuevo{
			In Figure \ref{FigKxvsKy}, the radial scale $k_x$ of the most unstable modes is shown versus the normalized $k_y$ wavenumber for the scan performed with EUTERPE in the W7-X configuration shown in Figure \ref{FigGRFR2}. The radial scale is obtained by Fourier transforming the potential profile for the most unstable mode at each single simulation in the scan. In order to check to what extent the radial scale $k_x$ is affected by the $\eta_i$ profile, 
			another simulation scan using a narrower  $\eta_i$ profile (see Figure \ref{FigProfiles}) was analyzed  and the results (shown in Figure \ref{FigKxvsKy}) are qualitatively the same in both cases. In 
			this scan the values obtained for $\gamma$ and $\omega$ were very close to those shown in Figure \ref{FigGRFR2} for the reference scan. }
		
		\nuevo{In spite of the differences, \nuevo{a remarkable consistency between results from the global and FT simulations is found, and} some important features are equally captured by both codes and computational domains. Firstly, for small wavenumbers $k_y\rho_r<2$ the radial scale of the most unstable mode in the RG domain is small $k_x\rho_r<0.5$,  in qualitative agreement with the results for the bean FT at short lengths, which shows the largest instability in this region of the maps. The radial wavenumbers of the unstable modes increase with $k_y$ up to $k_x \sim 1.5$ around $k_y \sim 2$. Interestingly, these radial scales $k_x\rho_r \sim 1.5$ and $k_y\rho_r \sim 2$ are the most unstable in the triangular FT  at $n_{pol}=1,2$ in the region $k_y\rho_r<2$, while for $n_{pol}<3$ they do not appear among the most unstable scales in the bean FT. For $2<k_y\rho_r<5$, the $k_x\rho_r$ of the most unstable mode decreases to very small values around $k_y\rho_r\sim4$,  which coincides qualitatively with the region in which the maximum growth rate is observed for small values of $k_x\rho_r$ in both FTs in the maps from \texttt{stella}. In this region $k_y\rho_r\sim 4$ a local minimum in the growth rate was shown in Figures \ref{FigGRFW7X-FTs} and \ref{FigGRFR2} which also appears in these maps.} \nuevo{Finally, for the region $k_y\rho_r>4$ an increase of $k_x\rho_r$ for the most unstable mode with $k_y\rho_r$ is obtained in the RG domain that seems to reflect the large spot in the maps from \texttt{stella}, particularly for the triangular FT, in which the maximum growth rate moves to regions of large values of $k_x\rho_r$ for large $k_y\rho_r$.
			  Given the differences between the FT and RG representations of the geometry, the consistency in results seems satisfactory. } 
		  
		  \nuevo{The FT and RG domains provide different perspectives of the same 3D reality. The flux tubes sample a reduced region of the flux surface and are local in radius, while the RG simulation takes all the 3D structure into account. Consequently, one can expect that modes that are very unstable in a FT geometry can become screened when they are allowed to explore the full magnetic structure.  The opposite situation can also be expected: modes that are not very unstable in a FT reduced domain can enhance its unstable properties when they explore larger regions of the flux surface/radius in the RG simulation. 
		  	This effect appears when the triangular flux tube is extended from 1 to 3 poloidal turns, and the growth rate obtained increases up to similar values to those of the bean FT. Then, the picture obtained from a global simulation represents something ``in between" the many FT that can be constructed in a flux surface. }

		\section{Summary and conclusions}\label{secSumandConcls}
   Two linear problems have been studied in gyrokinetic simulations carried out with different codes (GENE, GENE-3D, \texttt{stella} and EUTERPE) in different computational domains: flux tube, radially local full-flux-surface, and radially global domains.     
For the linear zonal flow relaxation problem, it has been shown that different flux tubes with the shortest length considered (one poloidal turn)  give different results, in general. The results of different FTs get closer as their length is increased. Both the length required for convergence of FT results and the degree of convergence are configuration-dependent, in line with previous results \cite{Smoniewski19}. In the particular case of LHD, different FTs give very similar results, which are also in reasonable agreement with RG \nuevo{and FFS} simulations, particularly for small radial wavenumbers. \nuevo{However, in the W7-X configuration analyzed here the situation is very different and different FTs with one or two poloidal turns long provide different results \nuevo{for the long-term properties (ZF residual and oscillation frequency).  For the ZF evolution at short times, no significant difference is found between FTs}.  }

\nuevo{Not only the long-term properties  but also the short-time evolution of the  zonal structures show a dependency with the radial scale of the perturbation, which is in very good agreement between simulations in long-enough FTs and a RG domain.}

Related to the ITG linear stability, the situation is somehow similar, but the growth rate of unstable modes captured by the FT local simulations is found to be dominated by the most unstable modes, with amplitude peaking at specific locations over the flux surfaces, which is in line with the strong localization of ITG (and also TEM) modes in stellarators found in previous works \cite{Nadeem01,Kornilov04,Xanthopoulos14b,Sanchez19}. 
In LHD, results of different flux tubes of any length match very well and are in good agreement with full surface or global simulation results. Radially global simulations are in very good agreement with FT and FFS in the frequency of the most unstable mode, but give a smaller (15-20\%) growth rate.
In W7-X, however, different flux tubes provide significantly different results {for $k_x=0$} that get closer {with each other} as the FT length is increased, but they do not exactly match at any length $n_{pol}\leq10$.

\nuevo{The radial scale of unstable ITG modes shows a dependency with the binormal wavenumber in W7-X configuration, while in LHD only a weak dependency with $k_y$ was found.  The dependency of the radial scale of the most unstable modes with $k_y$  in W7-X differs between FTs for short FT lengths. For long-enough flux tubes a qualitative consistency with radially global results  is found in this respect, in spite of the differences between both computational domains.}

The good agreement between different FTs, of any length, found in LHD for both the linear ITG instability, and for  the linear ZF evolution, cannot be considered a  situation valid for other devices, in general. The results for the W7-X configuration, as well as previous results in quasisymmetric devices \cite{Smoniewski19}, show that the convergence between results from different flux tubes is configuration-dependent.  Furthermore, 
the sampling of the flux surface by flux tubes of increasing lengths  is largely conditioned by the rotational transform. Besides, the situation for other different instabilities, such as trapped electron modes, which can be located preferentially at different locations than those where the ITG modes, can be very different. 

\nuevo{FFS and RG results are in reasonable agreement, which indicates that the FFS simulation domain can be considered the minimum appropriate computational domain for stellarators. However, slight differences between simulations in these domains appear.}

Work is in progress to study how the linear stability and zonal flow evolution studied here translate in simulations of nonlinearly saturated ITG turbulence in stellarators.

\section{\textbf{Acknowledgements}}
Simulations were carried out in Mare Nostrum IV and Marconi supercomputers. We acknowledge the computer resources and the technical support provided by the Barcelona Supercomputing Center and the EUROfusion infraestructure at CINECA. The work has been partially funded by the Ministerio de Ciencia, Innovación y Universidades of Spain under project PGC2018-095307-B-I00. This work has been carried out within the framework of the EUROfusion Consortium and has received funding from the Euratom research and training programme 2014-2018 and 2019-2020 under grant agreement N$^o$ 633053. The views and opinions expressed herein do not necessarily reflect those of the European Commission.


\end{document}